\documentstyle[psfig]{mn}

\newif\ifAMStwofonts

\ifoldfss
  \ifCUPmtlplainloaded \else
    \NewTextAlphabet{textbfit} {cmbxti10} {}
    \NewTextAlphabet{textbfss} {cmssbx10} {}
    \NewMathAlphabet{mathbfit} {cmbxti10} {} 
    \NewMathAlphabet{mathbfss} {cmssbx10} {} 
  \fi
  \ifAMStwofonts
    \ifCUPmtlplainloaded \else
      \NewSymbolFont{upmath} {eurm10}
      \NewSymbolFont{AMSa} {msam10}
      \NewMathSymbol{\upi}     {0}{upmath}{19}
      \NewMathSymbol{\umu}     {0}{upmath}{16}
      \NewMathSymbol{\upartial}{0}{upmath}{40}
      \NewMathSymbol{\leqslant}{3}{AMSa}{36}
      \NewMathSymbol{\geqslant}{3}{AMSa}{3E}

       \let\le=\leqslant
       
    \fi
  \fi
\fi 

\ifnfssone
  \newmathalphabet{\mathit}
  \addtoversion{normal}{\mathit}{cmr}{m}{it}
  \addtoversion{bold}{\mathit}{cmr}{bx}{it}
  \newmathalphabet{\mathbfit} 
  \addtoversion{normal}{\mathbfit}{cmr}{bx}{it}
  \addtoversion{bold}{\mathbfit}{cmr}{bx}{it}
  \newmathalphabet{\mathbfss} 
  \addtoversion{normal}{\mathbfss}{cmss}{bx}{n}
  \addtoversion{bold}{\mathbfss}{cmss}{bx}{n}
  \ifAMStwofonts
    \ifCUPmtlplainloaded \else
      \UseAMStwoboldmath
      \makeatletter
      \new@mathgroup\upmath@group
      \define@mathgroup\mv@normal\upmath@group{eur}{m}{n}
      \define@mathgroup\mv@bold\upmath@group{eur}{b}{n}
      \edef\UPM{\hexnumber\upmath@group}
      \new@mathgroup\amsa@group
      \define@mathgroup\mv@normal\amsa@group{msa}{m}{n}
      \define@mathgroup\mv@bold\amsa@group{msa}{m}{n}
      \edef\AMSa{\hexnumber\amsa@group}
      \makeatother
      \mathchardef\upi="0\UPM19
      \mathchardef\umu="0\UPM16
      \mathchardef\upartial="0\UPM40
      \mathchardef\leqslant="3\AMSa36
      \mathchardef\geqslant="3\AMSa3E

       \let\le=\leqslant

    \fi
  \fi
\fi 

\ifnfsstwo
  \DeclareMathAlphabet{\mathbfit}{OT1}{cmr}{bx}{it}
  \SetMathAlphabet\mathbfit{bold}{OT1}{cmr}{bx}{it}
  \DeclareMathAlphabet{\mathbfss}{OT1}{cmss}{bx}{n}
  \SetMathAlphabet\mathbfss{bold}{OT1}{cmss}{bx}{n}
  \ifAMStwofonts
    \ifCUPmtlplainloaded \else
      \DeclareSymbolFont{UPM}{U}{eur}{m}{n}
      \SetSymbolFont{UPM}{bold}{U}{eur}{b}{n}
      \DeclareSymbolFont{AMSa}{U}{msa}{m}{n}
      \DeclareMathSymbol{\upi}{0}{UPM}{"19}
      \DeclareMathSymbol{\umu}{0}{UPM}{"16}
      \DeclareMathSymbol{\upartial}{0}{UPM}{"40}
      \DeclareMathSymbol{\leqslant}{3}{AMSa}{"36}
      \DeclareMathSymbol{\geqslant}{3}{AMSa}{"3E}

       \let\le=\leqslant

    \fi
  \fi
\fi 

\ifCUPmtlplainloaded \else
  \ifAMStwofonts \else 
    \def\upi{\pi}
    \def\umu{\mu}
    \def\upartial{\partial}
  \fi
\fi

\title[Formation of the LMC]{Formation and evolution of the Magellanic 
Clouds. I.Origin of structural, kinematical, and chemical properties of 
the Large Magellanic Cloud}
\author[K. Bekki,  M. Chiba]
       {Kenji Bekki,${}^1$ and Masashi Chiba${}^2$ \\
        ${}^1$School of Physics, University of New South Wales, Sydney 2052, 
NSW, Australia \\
        ${}^2$Astronomical Institute, Tohoku University, Sendai, 
980-8578, Japan}
\date{Accepted 
      Received
      in original form 2001}

\pagerange{\pageref{firstpage}--\pageref{lastpage}}
\pubyear{1994}

\begin{document}

\maketitle

\label{firstpage}

\begin{abstract}

We investigate the dynamical and chemical evolution of the Large Magellanic
Cloud (LMC) interacting with the Galaxy and the Small Magellanic Cloud (SMC)
based on a series of self-consistent chemodynamical simulations.
Our numerical models are aimed at explaining the entire properties of the LMC,
i.e., the observed structure and kinematics of its stellar halo and disk
components as well as the populations of the field stars and star clusters.
The main results of the present simulations are summarized as follows.

(1) Tidal interaction between the Clouds and the Galaxy during the last 9 Gyr
transforms the initially thin, non-barred LMC disk into the three different
components: the central bar, thick disk, and kinematically hot stellar halo.
The central bar is composed both of old field stars and newly formed ones
with each fraction being equal in its innermost part.
The final thick disk has the central velocity dispersion of
$\sim$ 30 km s$^{-1}$ and shows rotationally supported kinematics 
with $V_{\rm m}/{\sigma}_{0}$  $\sim$ 2.3.

(2) The stellar halo is formed during the interaction, consisting mainly of
old stars originating from the outer part of the initially thin LMC disk.
The outer halo shows velocity dispersion of $\sim$ 40 km s$^{-1}$ 
at the distance of 7.5 kpc from the LMC center and has somewhat inhomogeneous
distribution of stars. The stellar halo contains relatively young,
metal-rich stars with the mass fraction of 2 \%.

(3) Repetitive interaction between the Clouds and the Galaxy enhances moderately
the star formation rate to $\sim$ 0.4 $M_{\odot}$ yr$^{-1}$ in the LMC disk.
Most of the new stars ($\sim$ 90 \%) are formed within the central 3 kpc of
the disk, in particular, within the central bar for the last 9 Gyr. 
Consequently, the half mass radius is different by a factor
of 2.3 between old field stars and newly formed ones.

(4) Efficient globular cluster formation does not occur until the LMC starts
interacting violently and closely with the SMC ($\sim$ 3 Gyrs ago).
The newly formed globular cluster system has a disky distribution with
rotational kinematics and its mean metallicity is $\sim$ 1.2 higher than that
of new field stars because of the pre-enrichment by the formation of
field stars prior to cluster formation.

(5) The LMC evolution depends on its initial mass and orbit with respect to
the Galaxy and the SMC. In particular, the epoch of the bar and thick disk
formation and the mass fraction of the stellar halo depend on the initial
mass of the LMC.

Based on these results, we discuss the entire formation history of
the  LMC,  the possible fossil records of past interaction between
the Clouds and the Galaxy, and the star formation history of
the SMC for the last several Gyr.
\end{abstract}


\section{Introduction}

Galaxy interaction is generally considered  to play a major role not only in
controlling formation histories of field stars and star clusters
(e.g., Kennicutt 1998; Ashman \& Zepf 1992) 
but also in transforming galactic morphologies (e.g., Noguchi 1987).
The Large and Small Magellanic Cloud (LMC and SMC),
which are believed to be interacting with each other, 
have long been served as an  ideal laboratory to study 
the detail of the tidal effects on 
structural, kinematical, and chemical properties  of galaxies 
based on the comparison between observations and numerical simulations
(e.g., Murai \& Fujimoto 1980, hereafter MF;
Gardiner, Sawa, \& Fujimoto 1994, GSF; Gardiner \& Noguchi 1996, GN;
Yoshizawa \& Noguchi 2003, YN).
Most of previous theoretical and numerical papers on
tidal interaction between the Clouds and the Galaxy however
have discussed  the origin of 
Magellanic stream and the evolution of the SMC
(e.g., Lin \& Lynden-Bell 1977, 1982; Mathewson et al. 1987; MF; GSF; GN; YN)
rather than the formation and the evolution of the LMC itself.
Given the fact that 
recent observations have raised and confirmed several important questions 
related to the star formation history,
the physical properties of star clusters, 
and the dynamical properties of the LMC,
it is doubtlessly worthwhile to discuss a theoretical model
which provides an integrated and systematic understanding of the formation 
and the evolution of the LMC (Westerlund 1997).

One of the long-standing and remarkable problems related to the star formation 
history of field stars in the LMC is as to whether there were
epochs of dramatic increase in star formation rate
of the LMC disk a few or several Gyr ago (e.g.,  Butcher 1977; Stryker 1983; 
Bertelli et al. 1992;  Olszewski et al. 1996; Gallagher et al. 1996;
Vallenari et al. 1996;  Holzman et al. 1999). 
Several authors have recently investigated
star formation histories of the different regions of the LMC disk and bar 
based on the color magnitude diagrams of the field stars
derived by {\it Hubble Space Telescope} 
(Ardeberg et al. 1997; Elston et al. 1997; 
Holzman et al. 1999; Olsen 1999; Smecker-Hane et al. 2002),
and revealed that the epoch of enhanced star formation 
and the degree of the enhancement  
are different between different regions.
For example, Smecker-Hane et al. (2002) found that
(1) stellar populations within the LMC bar were
formed in the episodes of  star formation about  $4-6$ 
and 1-2 Gyr ago, and (2) these burst populations can account
for $\sim$ 25 \% and $\sim$ 15 \% of the LMC's stellar mass, respectively.
The origin of the observed spatially different star formation
histories in the LMC disk is one of key questions related to the LMC evolution
(e.g., van den Bergh 2000a).

Several important physical properties of
the globular clusters and populous young blue clusters in the LMC
are in a stark contrast to those in the Galaxy (e.g., van den Bergh 2000a).
These include the more flattened shapes of the LMC clusters (e.g., Geisler \& Hodge 1980;
van den Bergh \& Morbey 1984), the disky distribution of its globular cluster 
system (e.g., Schommer et al. 1992),  possible rotational kinematics
of old clusters (e.g., Freeman et al. 1983), a larger fraction of apparently binary clusters
or physical cluster pairs in the LMC
(Bhatia \& Hatzidimitriou 1988; Bhatia et al. 1991; Dieball \& Grebel 1998),
a possible  ``age/metallicity  gap''
(e.g., Da Costa 1991; Olszewski et al. 1991;
Geisler et al. 1997; Sarajedini 1998; Rich et al. 2001),
and larger sizes at a given galactocentric distance (van den Bergh 2000b). 
It is unclear whether these differences are understood in terms of
the LMC being much more strongly influenced dynamically by other nearby galaxies
(i.e., the Galaxy and the SMC)
compared with the Galaxy.

Structural and kinematical properties of the LMC have been investigated
by many authors concerning different stellar populations and gaseous components
(Hartwick \& Cowley 1988; Meatheringham et al. 1988;
Irwin 1991; Luks \& Rohlfs 1992; Kunkel et al. 1997;
Graff et al. 2000; Olsen \& Salyk 2002; Cioni \& Habing 2003; 
Staveley-smith et al. 2003; Subramaniam 2004;  See Westerlund 1997 for a review).
For example, Caldwell \& Coulson (1986) found that the east side of the LMC
is closer to us than the west based on photometric observations of carbon stars.
Wide-field photometric observations of the LMC disk 
revealed the exponential scale length 
of 1.5 kpc (Bothun \& Thompson 1988) whereas an exponential disk
model that fits to the distribution of field RR Lyrae stars 
has a scale length of  2.6 kpc (Kinman et al. 1991).
Based on observational data of carbon stars by 
the Deep Near-Infrared Southern Sky Survey (DENIS)
and Two Micron All-Sky Survey (2MASS), 
van den Marel et al. (2002) (hereafter vdMAHS) have recently
shown that the LMC has a considerable vertical thickness
with $V/\sigma$ of 2.9$\pm$0.9 and 
a mass of (8.7$\pm$4.3)$\times$ $10^9$ $M_{\odot}$ within 8.9 kpc. 
The origin of the observed differences in dynamical properties between various
populations in the LMC remains unclear,
though such differences may well provide
some information about physical roles of galaxy interaction
in the LMC evolution (Westerlund 1997; van den Bergh 2000a).

Only some theoretical attempts have been made to understand these observed
properties of the LMC disk (Pagel et al. 1998; GSF; Weinberg 2000;
Kumai et al. 1993), while a growing number of observational results have emerged.
Weinberg (2000) first investigated the dynamical effects of the Galactic tidal field
on structure and kinematics of the LMC disk. Bekki et al. (2004a) demonstrated that
a single or binary star cluster is formed in a cloud-cloud collision
triggered by tidal interaction between the LMC and SMC. 
However, these studies have the following two
disadvantages in understanding the LMC evolution in a comprehensive manner. 
Firstly, these models are not fully self-consistent as they consider either
dynamical evolution alone or empirical one-zone chemical evolution. 
Secondly, the model parameters adopted in these works did not account for the
revised knowledge on the structure and kinematics of the LMC as revealed by
the latest observations such as DENIS and 2MASS
(van der Marel \& Cioni 2001; vdMAHS). Thus,
the integrated and systematic understanding of structural, kinematical, and
chemical properties of the LMC is therefore yet to be obtained.

Thus, the purpose of this paper is to investigate these unresolved problems
on the LMC formation and evolution based on chemodynamical simulations
with the model parameters (such as LMC's mass) consistent with 
the latest observations.
In particular, we examine the following five issues:
(1) how the star formation history of the LMC is influenced by tidal interaction
between the LMC, the SMC, and the Galaxy,
(2) whether the stellar halo formation process of the LMC is similar to or
different from that of other members of the Local Group of galaxies 
(e.g., M33, the Galaxy, NGC 3109, and NGC 6822),
(3) how we can explain the origin of structure and kinematics of the thin/thick
stellar disks of the LMC,
(4) how the LMC's chemical evolution process is associated with the dynamical
evolution influenced both by the Galaxy and the SMC,
and (5) whether the origin of the unique nature of the globular clusters and
blue populous clusters is understood in terms of the tidal interaction
between the Clouds and the Galaxy.
Based on these investigations, we attempt to provide a entire history of the LMC
for the last $\sim$ 10 Gyr.

The layout of this paper is as follows. In \S 2, we summarize our numerical
models used in the present study and describe the methods for analyzing
structure and kinematics of the simulated LMC. In \S 3, we present numerical
results on the time evolution of morphology, metallicity distribution, and
dynamical properties of the LMC. In \S 4, we discuss the above five outstanding
issues related to formation and evolution of the LMC. 
The conclusions of the preset study are given in \S 5.

\begin{table*}
\centering
\caption{Orbit models \label{tbl-1}}
\begin{tabular}{ccccc}
model &
$M_{\rm LMC}$ ($\times 10^{10}M_{\odot}$) & 
Dynamical friction & 
($U_{\rm L},V_{\rm L},W_{\rm L}$) (kms$^{-1}$) & 
($U_{\rm S},V_{\rm S},W_{\rm S}$) (kms$^{-1}$)  \\
A  & 1.0  & yes &  (-5,-225,194) & (40,-185,171)  \\
B  & 1.0  & no &  (-5,-225,194) & (40,-185,171)  \\
C  & 2.0  & yes &  (-5,-225,194) & (40,-185,171)  \\
D  & 2.0  & no &  (-5,-225,194) & (40,-185,171)  \\
E  & 1.0  & yes &  (-56,-219,186) & (60,-174,173)  \\
F  & 1.0  & yes &  (41,-200,169) & (60,-174,173)  \\
G  & 1.0  & yes &  (-15,-225,194) & (40,-185,171)  \\
H  & 1.0  & yes &  (5,-225,194) & (40,-185,171)  \\
I  & 1.0  & yes &  (-2,-225,194) & (40,-185,171)  \\
J  & 1.0  & yes &  (-5,-225,194) & (50,-185,171)  \\
K  & 1.0  & -- &  (-5,-225,194) & (40,-185,171)  \\
\end{tabular}
\end{table*}

\begin{figure}
\psfig{file=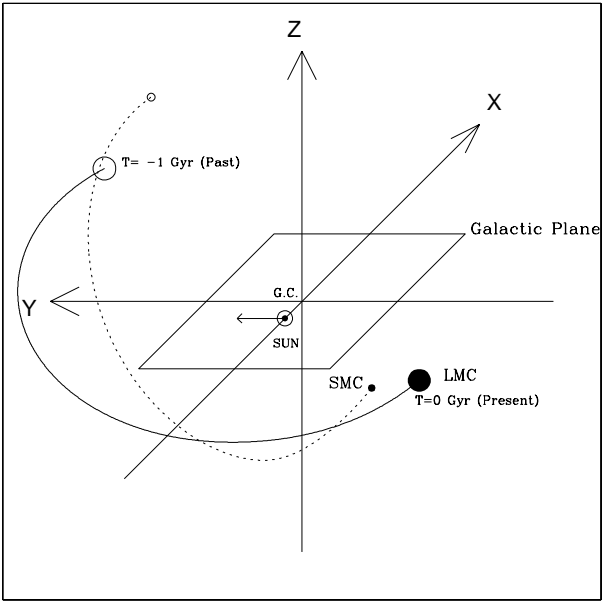,width=8.cm}
\caption{
Schematic view of the Galaxy and the Magellanic Clouds.
The present position of the LMC and that of the SMC
are indicated by a large filled circle and by
a small one, respectively, in the standard galactocentric
coordinate system ($X$, $Y$, $Z$).
The orbital evolution of the LMC and that of the SMC
for the last 1 Gyr ($-1$ $\le$ $T$ $\le$ 0 Gyr) are shown
by a solid line and by a dotted one, respectively.
}
\label{Figure. 1}
\end{figure}

\section{Model}

The present investigation is two-fold.
First, we derive the most plausible and realistic orbits of 
the Clouds with respect to the Galaxy 
by using a backward integration
scheme (MF; GSF; GN).
Then we investigate chemodynamical evolution of the LMC on the
derived orbit by using a fully self-consistent 
N-body model of galaxies (Bekki \& Shioya 1998, 1999;
Bekki \& Chiba 2000, 2001).
Since our main purpose of this paper is to discuss the origin of structural and kinematical
properties of the LMC, we briefly describe the results of the orbital evolution  derived from
the backward integration scheme in this section. 
We describe the results of the N-body simulations in the next section \S 3.

\subsection{Derivation of the orbits of the Clouds}

\subsubsection{The backward integration scheme}

We adopt the backward integration scheme originally devised by
MF to derive the most reasonable and 
realistic three dimensional (3D) orbits of the LMC and the SMC  
around the Galaxy. 
In calculating the orbital evolution of this triple interacting system,
we adopt the most recently derived observational parameters
on locations and radial velocities of the Clouds with respect to the Galaxy
and masses of the Clouds.
In order to derive the orbits of the Clouds,
we need to assume model parameters for the following quantities
of the Galaxy and the Clouds: (1) the shape of the Galactic
potential as a function of the distance $r$ from the Galactic center,
in particular, beyond 200 kpc where the Clouds reach at their
apocenter passages, 
(2) gravitational potential of the Clouds, 
(3) total masses (or mass profiles) of the Clouds,
(4) dynamical friction between the Galactic dark halo and the LMC (SMC),
and (5) dynamical friction between the LMC and the SMC.   
The above (3) and (5) are more carefully considered in the preset study
as described below.

The gravitational potential of the Galaxy ${\Phi}_{\rm G}$
is assumed to have the logarithmic potential;
\begin{equation}
{\Phi}_{\rm G}(r)=-{V_0}^2 \ln r ,
\end{equation}
where  $r$ and $V_{0}$ are the distance from the Galactic center
and the constant rotational velocity (= 220 km s$^{-1}$), respectively. 
The LMC is assumed to have the Plummer potential;
\begin{equation}
 {\Phi}_{\rm L}(r_{\rm L})=-M_{\rm LMC}/{({r_{\rm L}}^2+a_{\rm L}^2)}^{0.5},
\end{equation}
where $M_{\rm LMC}$, $r_{\rm L}$, and $a_{\rm L}$ are 
the total mass of the LMC, the distance from the LMC, and
the effective radius, respectively. We adopt the same value
of  $a_{\rm L}$ (=3 kpc) as previous numerical studies
adopted (e.g., GN).
Recent dynamical study of the LMC by vdMAHS
showed that the dynamical mass within 8.9 kpc of the LMC  
is (8.7 $\pm$ 4.3) $\times$ $10^9$ $M_{\odot}$,
which is less than the half of the mass (2.0 $\times$ $10^{10}$ $M_{\odot}$)
adopted in previous numerical studies (e.g., GN). 
Considering this latest and more robust estimation of the dynamical mass
of the LMC, we mainly investigate the orbital models with 
$M_{\rm LMC}$ = $10^{10}$ $M_{\odot}$.
The SMC is assumed to have the Plummer potential;
\begin{equation}
 {\Phi}_{\rm S}(r_{\rm L})=-GM_{\rm SMC}/{({r_{\rm S}}^2+a_{\rm S}^2)}^{0.5},
\end{equation}
where $M_{\rm SMC}$, $r_{\rm S}$, and $a_{\rm S}$ are 
the total mass of the SMC, the distance from the SMC, and
the effective radius, respectively. We adopt the same values
of  $a_{\rm L}$ (=3 kpc) and $M_{\rm SMC}$ 
(=3.0 $\times$ $10^{9}$ $M_{\odot}$) as previous numerical studies
adopted (e.g., GN).

 We consider the dynamical friction due to the presence of the Galactic
dark matter halo both for the LMC-Galaxy interaction
and for the SMC-Galaxy one and adopt the 
following expression (Binney \& Tremaine 1987);
\begin{equation}
F_{\rm fric, G}=-0.428\ln {\Lambda}_{\rm G} \frac{GM^2}{r^2},
\end{equation}
where $r$ is the distance of the LMC (the SMC)
from the center of the Galaxy.
The mass $M$ is either $M_{\rm LMC}$ or $M_{\rm SMC}$,
depending on which Cloud's orbit (i.e., LMC or SMC) we calculate.
We adopt the reasonable value of 3.0 
for the Coulomb logarithm ${\Lambda}_{\rm G}$
(GSF; GN)
both in  the orbital calculation of the LMC and in that of the SMC. 
The above equation  (4) is essentially the same as that shown in 
the equation (18) by MF. 
In addition to the above $F_{\rm fric, G}$,
the dynamical friction between the LMC and the SMC is also considered
in the present study. 
We adopt the following expression;
\begin{equation}
F_{\rm fric, LS}=-0.428\ln {\Lambda}_{\rm LS} \frac{G{M_{\rm SMC}}^2}{{r_{\rm LS}}^2},
\end{equation}
where $r_{\rm LS}$ and ${\Lambda}_{\rm LS}$
are the distance between the center of the LMC and that of the SMC
and the Coulomb logarithm, respectively.
$F_{\rm fric, LS}$ is assumed to act on the SMC,
only when the SMC is within the LMC's tidal radius $r_{\rm t}$  
within which dark matter halo is gravitationally 
bound without being stripped from
the Galactic tidal field.
By using the theoretical model  adopted in the equation (4) of GN,
$r_{\rm t}$ is estimated as 13 kpc for the present model.
vdMAHS observationally 
estimated $r_{\rm t}$ as 15.0 $\pm$ 4.5 kpc
based on the newly derived total mass of the LMC. 
Therefore our choice of  $r_{\rm t}$  = 13 kpc is regarded as
a quite reasonable value. 
For comparison, we investigate the models with  $r_{\rm t}$  =  0 kpc,
in which dynamical friction between the Clouds is not included at all.

By integrating equations of the motions of the Clouds
toward the past from the present epoch,
we investigate orbital evolution of the Clouds
for given initial positions and velocities of the Clouds.
We adopt the reasonable sets of orbital parameters that are 
consistent with observations and thus were adopted
in previous numerical studies (GN). 
Figure 1 shows a schematic view of the Galaxy and the Magellanic Clouds 
and the orbital evolution of the Clouds.
The current Galactic coordinate $(b,l)$, where $l$ and $b$ are
the Galactic longitude and latitude, respectively, is $(-32.89, 280.46)$
for the LMC and $(-44.30,302.79)$ for the SMC, and accordingly
the current positions $(X,Y,Z)$ in units of kpc in the figure
are $(-1.0,-40.8,-26.8)$  for the LMC
and $(13.6,-34.3,-39.8)$ for the SMC.
The current distance and the Galactocentric radial velocity of the LMC (SMC)
is 80 (7) km s$^{-1}$.

We try to investigate the LMC evolution between the epoch immediately after
its thin disk formation and the present-day. Oswalt et al. (1996) estimated 
the oldest stellar components of the Galactic thin disk as 9.5 $\pm$ 1 Gyr
based on the faint end of luminosity function of white dwarfs in the Galaxy.
There are no observational studies that
are based on the luminosity function of the white dwarfs of the LMC disk
and thus reveal how old the oldest stellar components
of the LMC disk are. 
Therefore, as a compromise,
we adopt the assumption that
the age of the oldest stellar populations of the LMC
is similar to that of the oldest ones in the Galaxy:
We thus investigate the LMC evolution for 
the last $\sim$ 9 Gyr.

\begin{figure}
\psfig{file=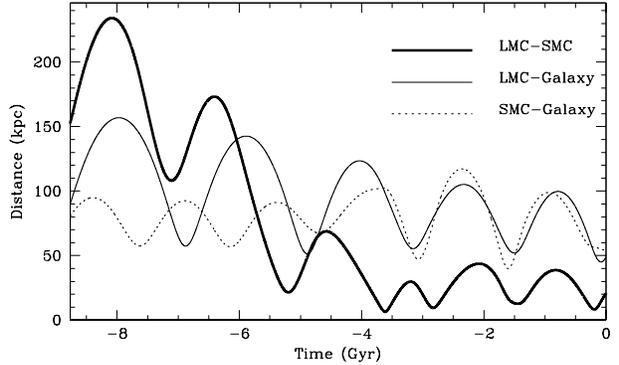,width=8.cm}
\caption{
The orbital evolution of the Clouds for the last $\sim$ 9 Gyr
($-8.8$ $\le$ $T$ $\le$ 0 Gyr) for the best orbital model A.
The distance between the Cloud,
that between the Galaxy and the LMC,
and that between the Galaxy and the SMC
are represented by
a thick solid line, a thin solid one, and a dotted one,
respectively.
}
\label{Figure. 2}
\end{figure}

\begin{figure}
\psfig{file=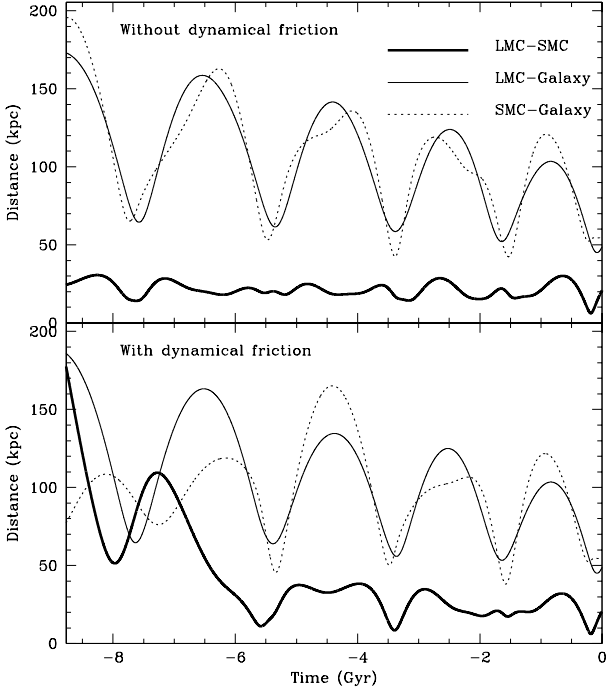,width=8.cm}
\caption{
The same as Figure 2 but for the orbital model C (lower)
and D (upper). It is clear from this figures that
the Clouds can not be in a binary state more than $\sim$ 6 Gyr
if the dynamical friction of the Clouds is included in
the orbital calculation.
}
\label{Figure. 3}
\end{figure}

\begin{figure}
\psfig{file=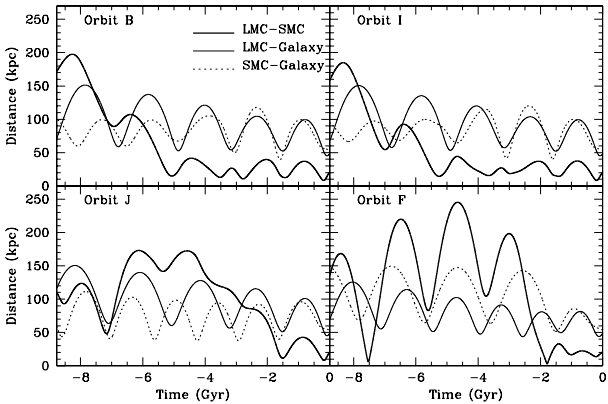,width=8.cm}
\caption{
The same as Figure 2 but for the orbital model B (upper left),
I (upper right),  J (lower left), and F (lower right).
The values of the model parameters for these orbits
are shown in the Table 1.
}
\label{Figure. 4}
\end{figure}

\subsubsection{The representative orbits}

The current space velocities 
or  $(U,V,W)$ in units of km s$^{-1}$ are the most important
parameters that determine the orbital evolution of the Clouds in the present models.
They  are represented by $(U_{\rm L},V_{\rm L},W_{\rm L})$
for the LMC and  by $(U_{\rm S},V_{\rm S},W_{\rm S})$ for the SMC.
Eleven representative models with different values of $(U_{\rm L},V_{\rm L},W_{\rm L})$
and $(U_{\rm S},V_{\rm S},W_{\rm S})$ are discussed  in the present study,
and  the Table 1 summarizes the model parameters for these:
Model number (column 1), 
total mass of the LMC  represented by $M_{\rm LMC}$ in units of $10^{10} M_{\odot}$ (2), 
whether or not the dynamical friction between the Clouds is included (3),
$(U_{\rm L},V_{\rm L},W_{\rm L})$ (4),
and $(U_{\rm S},V_{\rm S},W_{\rm S})$ (5).

Among these 11 models labeled as A$-$K, 
the model A 
with $(U_{\rm L},V_{\rm L},W_{\rm L})$ = $(-5,-225,194)$
and $(U_{\rm S},V_{\rm S},W_{\rm S})$  = $(40,-185,171)$
is considered to be the best orbital model and thus 
referred to as the ``best orbital model''
throughout this paper.  
This is 
mainly  because the Magellanic stream is self-consistently
reproduced in previous models for these values.
Different observations suggested different values of $(U_{\rm L},V_{\rm L},W_{\rm L})$
and $(U_{\rm S},V_{\rm S},W_{\rm S})$ (e.g., Kroupa \&  Bastian 1997, KB; vdMAHS).
For example, $(U_{\rm L},V_{\rm L},W_{\rm L})$ and 
$(U_{\rm S},V_{\rm S},W_{\rm S})$ are 
(41$\pm$44, $-200\pm$31, 169$\pm$37) km s$^{-1}$
and (60$\pm$172, $-174\pm$172, 173$\pm$128) km s$^{-1}$,
respectively,  for KB.
$(U_{\rm L},V_{\rm L},W_{\rm L})$ in vdMAHS
is ($-56\pm36$, $-219\pm23$, 186$\pm$35).
The best model accordingly is broadly consistent with
the above latest proper motion data (KB).
Taking this difference into account,
we investigate the orbital models with the initial velocity components
adopted from the above observations: Model E from vdMAHS,
model F from KB. We also investigate the orbital models with
the initial velocity components slightly different from those in the best model A:
Model G with $U_{\rm L}$ smaller by 10 km s$^{-1}$ than  that of the best model,
model H with $U_{\rm L}$ larger  by 10 km s$^{-1}$ than  that of the best model,
model I with $U_{\rm L}$ larger  by 2 km s$^{-1}$ than  that of the best model,
and model J with $U_{\rm S}$ larger  by 10 km s$^{-1}$ than  that of the best model.
To investigate the dynamical effects of the SMC on the orbital evolution of the LMC,
we investigate the model K without the SMC.

Figure 2 describes the past $\sim$ 9 Gyr orbital evolution of the  Clouds
in the best orbital model A.
Here  negative values of the time, $T$, represent the past, with
$T=0$ corresponding to the present epoch. 
As shown in this figure,
the present orbital period of the Clouds
about the Galaxy is $\sim$ 1.5 Gyr for the adopted
gravitational potential and the masses of the Cloud and the Galaxy in
the best model.  
Although the LMC-SMC distance
remains very small ($<40$\,kpc) over the last 4\,Gyr ($T >$  $-4$ Gyr),
it cannot keep its binary status more than $\sim$ 5 Gyr:
Disintegration of the present-day binary
orbit is inevitable in this model.

Figure 3 describes the orbital evolution of the Clouds
for the models C and D and thereby
shows how the dynamical friction between the Clouds due to the presence
of the dark halo of the LMC influences the orbital evolution of the Clouds.
The model D with  $M_{\rm LMC}$ = 2.0 $\times$ $10^{10}$ $M_{\odot}$
and without dynamical friction between the Clouds
is exactly the same as the best model in GN.
As shown in this figure, the Clouds can keep its binary status more than
$\sim$ 9 Gyr in the model D {\it  without  dynamical friction of the Clouds}
whereas they cannot
in the model C {\it  with dynamical friction}.
This suggests that the models with dynamical friction between the Clouds
have difficulties in keeping the binary status  of the Clouds
for more than several Gyrs.
In the Appendix A, we discuss the duration of the LMC/SMC binary status
more extensively based on large number of orbital models.
Figure 4 summarizes the orbital evolution 
for the representative models, B, I, J, and F,
for which we investigate the LMC evolution in the N-body simulations.

\begin{table*}
\centering
\caption{Model parameters for N-body simulations\label{tbl-2}}
\begin{tabular}{ccccc}
model no. &
orbit type & 
$\theta$ (degrees) & 
$\phi$ (degrees) & 
comments \\
1  & A &  99 & 257 &   fiducial  \\
2  & -- &  0 & 0 &   isolated LMC  \\
3  & A &  90 & 270 &   \\
4  & A &  79 & 257 &   \\
5  & A &  109& 257 &   \\
6  & A &  99 & 257 &  NFW halo\\
7  & A &  90 & 270 &  NFW halo\\
8  & B &  99 & 257 & \\
9  & C &  99 & 257 &   more massive LMC\\
10  & D &  99 & 257 &   more massive LMC\\
11  & E &  99 & 257 &   \\
12  & F &  99 & 257 &  \\
13  & G &  99 & 257 &   \\
14  & H &  99 & 257 &  \\
15  & I &  99 & 257 &   \\
16  & J &  99 & 257 &   \\
17  & K &  99 & 257 &  no SMC\\
18  & A &  99 & 257 & $f_{\rm r}$=0.0 \\
19  & A &  99 & 257 &  $f_{\rm r}$=0.25 \\
20  & A &  99 & 257 &  $f_{\rm r}$=0.75 \\
21  & A &  99 & 257 &  $f_{\rm r}$=1.0 \\
\end{tabular}
\end{table*}

\subsection{N-body models}

\subsubsection{The self-gravitating LMC}

The LMC is modeled as a fully self-gravitating system
and  composed of a live dark halo
and a thin exponential disk with no bulge.
The total mass of the dark halo,
that of the disk,  and the size of the disk  are $M_{\rm dm}$, $M_{\rm d}$
and $R_{\rm d}$, respectively. 
The mass ratio of the dark halo 
to the total mass (i.e., $M_{\rm dm}/M_{\rm LMC}$) is
fixed at 0.7 throughout the paper, which is consistent with
the observation by vdMAHS.
We adopt the Plummer potential in calculating the orbital evolution
of the LMC (\S 2.1.1) and  accordingly we need to adopt  the corresponding
density profile for the dark matter halo of the LMC
in this self-gravitating N-body models for self-consistency.
The density profile of the dark halo is described as:
\begin{equation}
{\rho}(r_{\rm L})=(\frac{3M_{\rm dm}}{4\pi {a_{\rm L}}^3}) 
{(1+\frac{{r_{\rm L}}^2}{{a_{\rm L}}^2})}^{-2.5},
\end{equation} 
where $r_{\rm L}$ is the distance from the center of the LMC
and the value of $a_{\rm L}$ is identical with that adopted in the equation
(2).
Recent cosmological simulations within the framework of the  CDM model  
 (Navarro, Frenk \& White 1996)
have demonstrated the ``universal'' density distribution 
(the NFW profile); 
 \begin{equation}
 {\rho}(r)=\frac{\rho_{0}}{(r/r_{\rm NFW})(1+r/r_{\rm NFW})^2},
 \end{equation} 
 where  $r$, $\rho_{0}$, and $r_{\rm NFW}$ are
the spherical radius,  the central density of a dark halo,  and the scale
length of the halo, respectively.  
For comparison, we also investigate the models with this NFW profile
for the LMC's dark matter halo, though it is not
totally self-consistent with our orbital models with the Plummer potential. 

The radial ($R$) and vertical ($Z$) density profile 
of the initially thin  disk of the LMC are  assumed to be
proportional to $\exp (-R/R_{0}) $ with scale length $R_{0}$ = 2.6 kpc 
and to  ${\rm sech}^2 (Z/Z_{0})$ with scale length $Z_{0}$ = $0.2R_{0}$, 
respectively.
We adopt the value of 2.6 kpc from Kinman et al. (1991) who 
investigated spatial distribution of RR Lyrae filed stars that are
believed to be old,
because our initial stellar disk is assumed to be older than  $\sim$ 9 Gyr. 
The circular velocity of the disk
becomes a maximum value of $V_{\rm m}$ 
at $r_{\rm L}$ =  5 kpc from the center of the disk  for the adopted dark matter
mass profile and $V_{\rm m}$ is 71 km s$^{-1}$.
In addition to the rotational velocity made by the gravitational
field of disk and halo component, the initial radial and azimuthal velocity
dispersion are given to the disk component according
to the epicyclic theory with Toomre's parameter (Binney \& Tremaine 1987) $Q$ = 1.5.
The vertical velocity dispersion at a given radius 
is set to be 0.5 times as large as
the radial velocity dispersion at that point, 
as is consistent  with 
the observed trend  of the Galaxy (e.g., Wielen 1977).
In order to compare physical properties of new GCs formed from gas
with those of old GCs initially within the LMC disk,
we assume that the LMC disk initially contains 100 old GCs 
and the GC system has a disky distribution and rotational kinematics.
We adopt the GC number  of 100 that is much larger than the observed one 
($\sim$ 13; van den Bergh 2000a),
because we need an order of 100 GCs to evaluate structural and kinematical
properties of the GC system: Only $\sim$ 10 GCs do not allow us to derive
the density profile and the rotational properties of the GC system.
The assumption on the GC kinematics and structure
is  consistent with observations of structural and
kinematics of old GCs in the LMC  (e.g., Freeman et al. 1983; Schommer et al. 1992).

 The disk is composed both of gas and stars with the gas mass fraction
($f_{\rm g}$) being a free parameter and the gas disk
is represented by a collection of discrete gas clouds 
(corresponding to giant molecular clouds; GMCs) that follow the observed mass-size
relationship (Larson 1981).
Every pair of two overlapping gas clouds
is made to collide with the same restitution coefficient of $f_{\rm r}$ 
(Hausman \& Roberts 1984).
We vary the values of  $f_{\rm r}$ from  $0.0$ (no dissipation)
1.0 (highly dissipative) and thereby investigate the parameter dependences
of the results on  $f_{\rm r}$. 
We mainly present the results of the models with  $f_{\rm r}$ =0.5
and show some parameter dependences of the present results on  $f_{\rm r}$.
Although  adopted method of ``sticky particles'' has been
proven to be capable of addressing successfully {\it some} aspects
of the hydrodynamical interactions in interstellar medium (ISM)  for disk galaxies
(e.g., Hausman \& Roberts 1984; Combes \& Gerin 1985),
it has some disadvantages in dealing with the more realistic
physical processes of the ISM, such as hydrodynamical interaction
between the hot interstellar gas (with the temperature of $10^6$ K)
and GMCs (e.g., evaporation of GMCs by the hot gas). 
We discuss this point in \S 3 for the results that may depend
sensitively on the way to treat with ISM.

In order to construct as a realistic gas disk model as possible,
we consider the radial dependence of the gas mass fraction 
$F_{\rm g}(r_{\rm L})$ in the initial LMC disk.
We expect that the inner gas mass fraction in the LMC is
smaller than the outer one owing to more rapid consumption of gas
in the inner regions with higher gas density.
We therefore adopt the following rule;
\begin{equation}
F_{\rm g}(r_{\rm L}) \propto t_{sf} (r_{\rm L})
\propto  \frac{{\Sigma}_{\rm g}(r_{\rm L})}
{\dot{\Sigma}_{\rm g}(r_{\rm L})}
 \propto {{\Sigma}_{\rm g}}^{\alpha} (r_{\rm L}),
\end{equation}
where $r_{\rm L}$,  $t_{sf}$, ${\Sigma}_{\rm g}$, 
$\dot{\Sigma}_{\rm g}$, and ${\alpha}$ are 
the distance from the center of the LMC disk, the gas consumption
time scale, the initial gas density,
the gas consumption rate, and the parameter controlling the radial dependence.
Since we adopt the Schmidt law with the  exponent of  1.5  
for star formation (described below),
the reasonable value of ${\alpha}$ is $-0.5$.
According to the value of $F_{\rm g}(r_{\rm L})$ 
derived from the above equation (8) at each radius,
we determine the reasonable number of old stellar particles and gaseous ones
at each radius and thereby allocate these particles to each radial bin. 
By assuming that the disk is composed only of gas initially,
we determine $F_{\rm g}(r_{\rm L})$ through the equation (8)
and derive a reasonable radial distribution  of gas and stars.
To obtain  more realistic {\it initial stellar and gaseous} distributions
(e.g., with gaseous and stellar spiral arms),
a LMC disk with $F_{\rm g}(r_{\rm L})$
is allowed to relax for 10 dynamical time  ($\sim$ 0.7 Gyr).
We then use this disk as an initial LMC disk model in the simulations. 

\subsubsection{Star formation and chemical evolution}

The gas is converted into either {\it field stars} or {\it globular clusters (GCs)},
so that we distinguish the formation process of field stars from that of GCs
throughout this paper.
Field star formation
is modeled by converting  the collisional
gas particles
into  collisionless new stellar particles according to the algorithm
of star formation  described below.
We adopt the Schmidt law (Schmidt 1959)
with exponent $\gamma$ = 1.5 (1.0  $ < $  $\gamma$
$ < $ 2.0, Kennicutt 1998) as the controlling
parameter of the rate of star formation.
The amount of gas 
consumed by star formation for each gas particle
in each time step 
is given as:
\begin{equation}
\dot{{\rho}_{\rm g}} \propto  
{\rho_{\rm g}}^{\gamma},
\end{equation}
where $\rho_{\rm g}$ 
is the gas density around each gas particle. 
We convert a gas particle into a field star only if the local
gas density ${\rho}$ exceeds the observed threshold gas density
of $\sim$ 3 $M_{\odot}$ pc$^{-2}$ for 
Magellanic  dwarf irregular galaxies (Hunter et al. 1998).
These stars formed from gas are called ``new stars'' (or ``young stars'')
whereas stars initially within a disk  are called ``old stars''
throughout this paper.

We use the cluster formation criteria derived by
previous analytical works (e.g., Kumai et al. 1993)
and hydrodynamical simulations  with variously different parameters of
cloud-cloud collisions on a 1-100pc scale (Bekki et al. 2004a) 
in order to model GC formation. A gas particle is converted into a cluster if it
collides with other high velocity gas (with the relative velocities
ranging from 30\,km\,s$^{-1}$
to 100\,km\,s$^{-1}$) and having an impact parameter (normalized to the
cloud radius) less than 0.25.
Although both binary cluster and single one are formed during
high-velocity cloud-cloud collisions (Bekki et al. 2004a),
we assume that only one cluster is formed from one event 
of cloud-cloud collision for simplicity.
These GCs formed from gas are called ``new GCs''
whereas GCs initially within a disk  are called ``old GCs''
throughout this paper.

This model is strongly supported by recent observations
(e.g., Zhang et al. 2001) that have revealed that
there is a tendency for young clusters to be found
in gaseous regions with higher velocity dispersion,
where cloud-cloud collisions are highly likely.
In the present model, the clusters are not formed
in the isolated model (described later) at all.
This enables us to investigate how and whether 
the tidal interaction between the Clouds and the Galaxy
triggers the formation of GCs.
About an order of $10^2$ clusters are formed in the models
where the LMC interacts with the SMC and the Galaxy.
The preset simulations  only investigate 
dynamics with the scale down to
$\sim$ 100 pc so that it does not allow us to investigate
which new clusters with the size of $\sim$ 10 pc
survive from tidal force of the Galaxy 
and the Clouds to be observed as GCs at the present time. 
Therefore, we assume  that 
{\it all clusters} formed in the simulations
become GCs,
and thereby analyze the physical properties of new GCs. 
The total number of GCs in the simulations may be 
overestimated in the present model because of this assumption.

 Chemical enrichment through star formation and supernovae feedback
 during the LMC evolution 
is assumed to proceed both locally and instantaneously in the present study.
We assign the metallicity of original
gas particle to  the new stellar particle and increase 
the metals of the each neighbor gas particle 
with the total number of neighbor gas particles equal to  $N_{\rm gas}$,
according to the following 
equation about the chemical enrichment:
  \begin{equation}
  \Delta M_{\rm Z} = \{ Z_{i}R_{\rm met}m_{\rm s}+(1.0-R_{\rm met})
 (1.0-Z_{i})m_{\rm s}y_{\rm met} \}/N_{\rm gas} 
  \end{equation}
where the $\Delta M_{\rm Z}$ represents the increase of metal for each
gas particle. $ Z_{i}$, $R_{\rm met}$, $m_{\rm s}$,
and $y_{\rm met}$  in the above equation represent
the metallicity of the new stellar particle (or that of original gas particle),
the fraction of gas returned to interstellar medium,  the
mass of the new star, and the chemical yield, respectively.
The values of $R_{\rm met}$,  $y_{\rm met}$, and the initial metallicity  are set to
be 0.3 and 0.005, and 0.002, respectively.
For these values, the final mean metallicity of
the inner region of a LMC disk 
is consistent with the observed one with [Fe/H] = $-0.3$ $\pm$ 0.04 (Luck et al. 1998).

In order to discuss the metallicity distribution of old stars stripped from the LMC disk,
we need to assume that the {\it old stellar} disk has a metallicity gradient 
consistent with observations.
Friel (1995) has derived  the metallicity gradient of the Galactic stellar disk 
based on the ages and metallicities that are estimated for the Galactic open
clusters. 
Since we do not have any available data on metallicity gradient of open clusters
in the LMC, we compromise to use the observed slope of the Galactic metallicity
gradient by Friel (1995) for the LMC old stellar disk.
We therefore  allocate metallicity to each disc star according to its initial position:
at $r_{\rm L}$ = $R_{\rm L}$,
where $r_{\rm L}$ ($R_{\rm L}$) is the projected distance in units of kpc
from the center of the LMC disk, the metallicity of the star is given as:
\begin{equation}
{\rm [m/H]}_{\rm r_{\rm L}=R_{\rm L}} = {\rm [m/H]}_{\rm d, r_{\rm L}=0} 
+ {\alpha}_{\rm d} \times {\rm R_{\rm L}}. \;
\end{equation}
If we adopt a  plausible values of $-0.091$ 
for the slope ${\alpha}_{\rm d}$ (Friel 1995)
and the central value of $-0.73$ for ${\rm [m/H]}_{\rm d, r_{\rm L}=0}$,
the mean metallicity of the LMC old disk is $-1.0$ in [Fe/H].

\begin{figure*}
\psfig{file=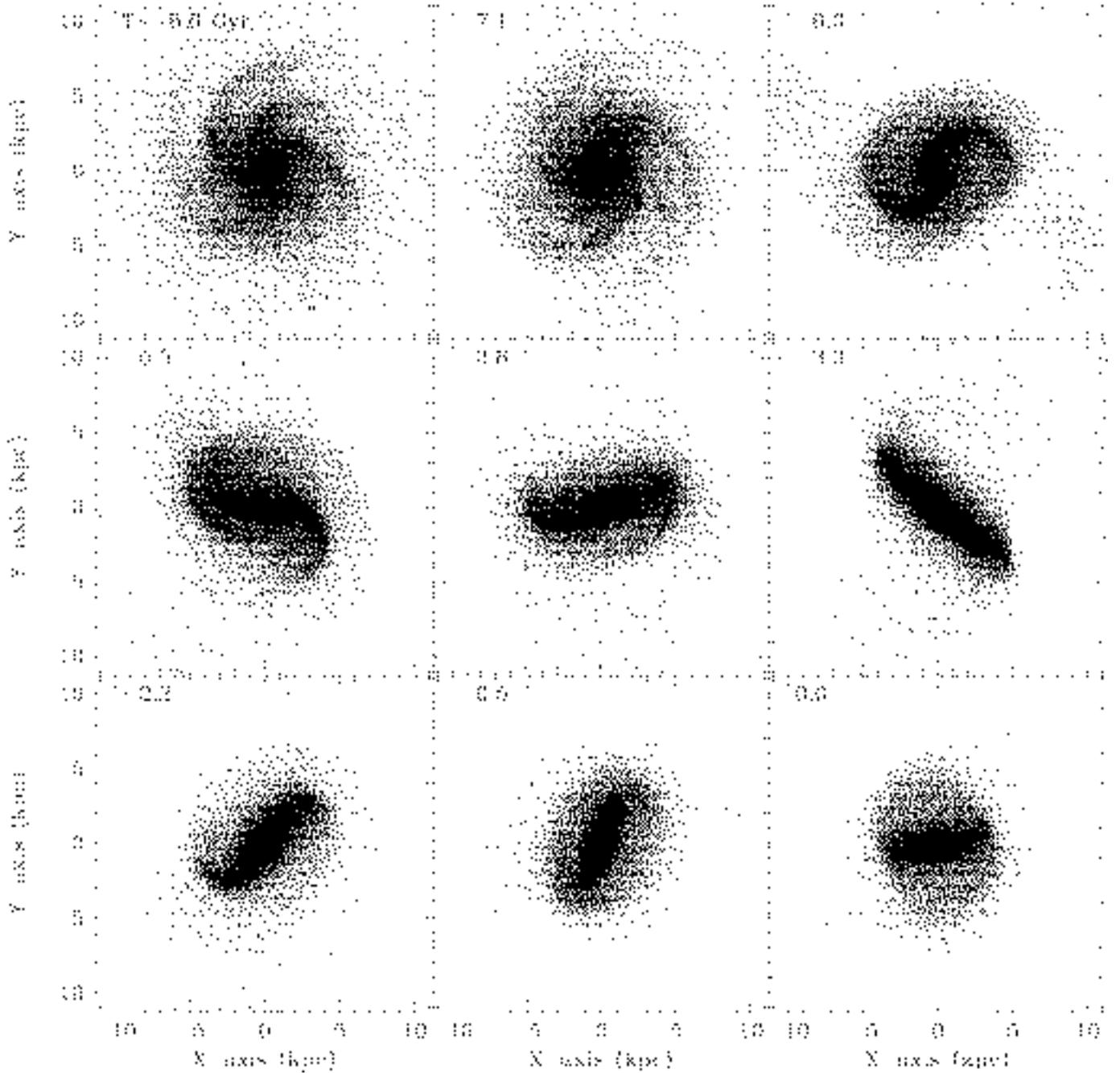}
\caption{
Morphological evolution of old stars and gas seen from the face-on view
(referred to as the $x$-$y$ plane)
with respect to the LMC disk
for the fiducial model.
}
\label{Figure. 5}
\end{figure*}

\subsection{N-body evolution on the pre-determined orbits}

We numerically investigate the evolution of the LMC disk 
under the gravitational influence by the Galaxy and the SMC
for the last $\sim$ 9 Gyr (128 dynamical time scales of the LMC) 
by using the above N-body models of the LMC. 
We adopt an orbital model (e.g., A) and consider that
the LMC (the SMC) in a simulation is always on the predetermined orbit. 
Based on the orbits of the Clouds  derived in the above 
orbital calculations (i.e., models A $-$ K),
we create a look-up table of positions and velocities of the Clouds
at each time step 
for $-8.8$ $\le$  $T$  $\le$ 0 Gyr with the time step width
of 1.4 $\times$ 10$^6$ yr
in each orbital model. 
The center of mass in the LMC (the SMC) at each time step in a simulation
with an orbital model
is set to be the same as the location
of the LMC (the SMC) at the time step in the look-up table of
the orbital model. 
A fully self-gravitating LMC model is influenced by the 
same fixed potential of the Galaxy and the SMC  used in 
the equation (1) and (3), respectively, in a simulation.
Since we do not intend to 
constitute the SMC as a self-gravitating particle system 
in the present study,
possible important physical processes such
as direct hydrodynamical interaction between gaseous components
of the Clouds 
and mass-transfer from the SMC to the LMC are not included.

\begin{figure*}
\psfig{file=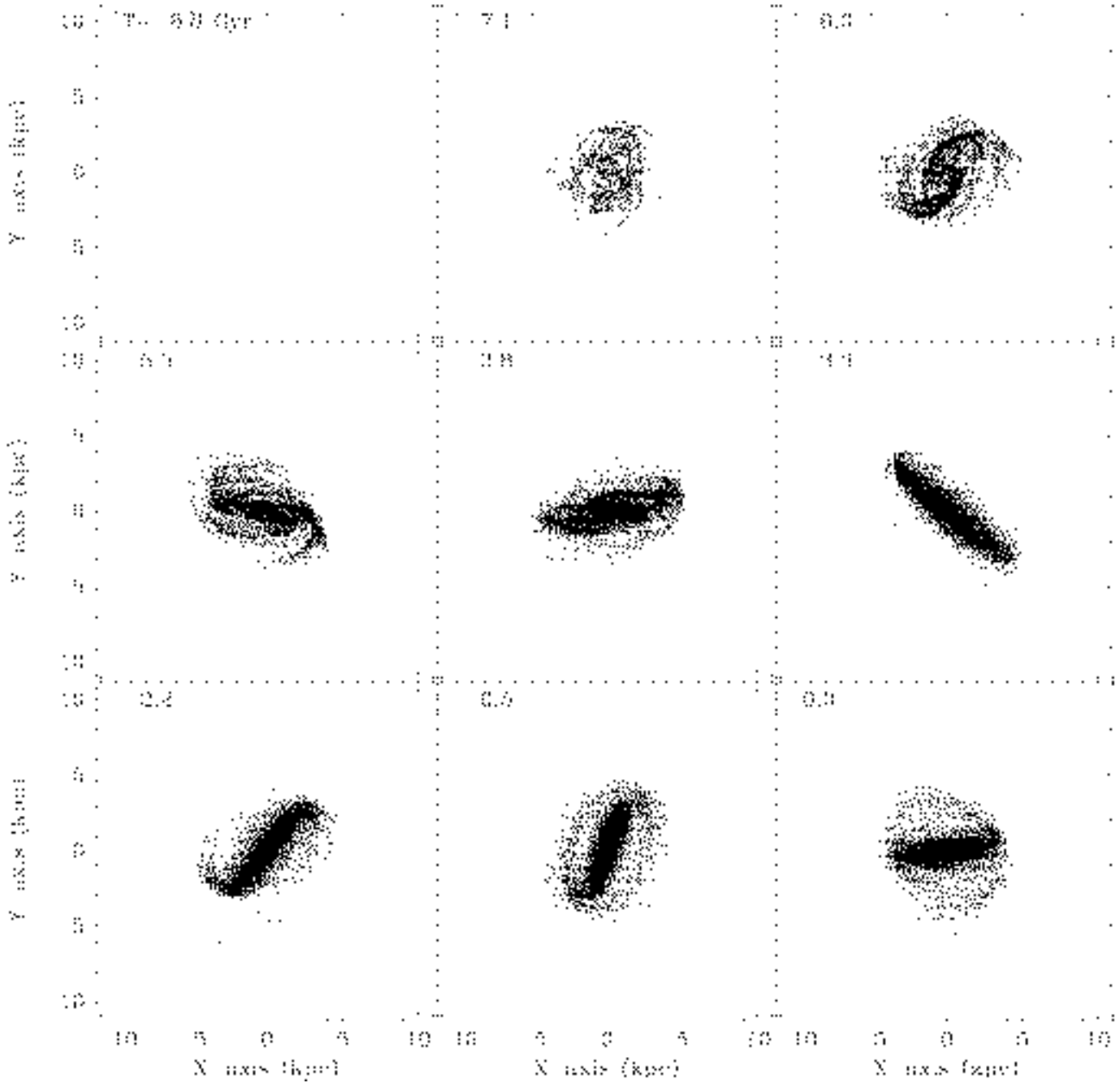}
\caption{
The same as Figure 5 but for new stars formed from gas.
}
\label{Figure. 6}
\end{figure*}

The initial spin of a LMC disk in  a model is specified by two angles,
$\theta$ and $\phi$, where
$\theta$ is the angle between the $Z$-axis and the vector of
the angular momentum of a disk and
$\phi$ is the azimuthal angle measured from $X$-axis to
the projection of the angular momentum vector 
of a disk onto the $X-Y$ plane.
We mainly describe the results of  the models with  
$\theta$ = 99$^{\circ}$ and $\phi$  = 257$^{\circ}$,
because the final structural properties
 of the LMC disk at $T$ = 0 Gyr in this model
is broadly  consistent with observations in that 
(1) the simulated LMC disk has a stellar bar with the size of the bar similar to
the observed one,
(2) the position angle of the simulated bar projected onto the sky   
is not largely different from  the observed one (yet not exactly the same),
and (3) the northeast side of the simulated disk is the near side with respect
to the Galaxy, which is consistent with the observations by Caldwell \& Coulson (1986).
The LMC disk precesses and nutates  under the tidal torque from
the Galaxy during the dynamical evolution of the LMC (Weinberg 2000). 
Therefore it is very hard to choose the {\it initial values} of $\theta$ and $\phi$
for which the {\it final} internal spin  axis 
of the LMC at $T$ = 0 in our simulations 
is nearly the same as those inferred from observations.
Thus,  we show the results of the models for which the final structural properties
of the LMC are broadly consistent with observations.

We mainly describe the ``fiducial model'' with the best orbital model/type A,
$\theta$ = 99$^{\circ}$, $\phi$ = 257$^{\circ}$, $f_{\rm g}$ = 0.5,
and $f_{\rm r}$ = 0.5. 
This is firstly because the parameters of this model are 
the most reasonably consistent with observations,
and  secondly because this model shows typical behaviors
in dynamical and chemical evolution of the LMC influenced
by the Galaxy and the SMC.
We also show the results of the representative 21 models thereby
discuss  the dependences of the results
on model parameters such as
the orbital types, $\theta$,  $\phi$, $f_{\rm g}$, and  $f_{\rm r}$.
The values of these parameters are summarized for each model in the Table 2:
Model number (column 1), orbital type (2),  $\theta$ in units of degrees (3),
$\phi$ in units of degrees (4), and the comments on the models (5).

The initial particle number used in a self-gravitating 
LMC  model  is 50000 for the dark matter,
25000 for the old stars, and 25000 for the gas.
The total particle number is increased up to $\sim$ 160000
owing to the formation of new field stars and globular clusters.
All the simulations have been carried out on
GRAPE board (Sugimoto et al. 1990) 
at the Astronomical Data Analysis Center (ADAC)
of the National Astronomical Observatory of Japan.
The parameter of gravitational softening is set to be fixed at 0.15 kpc. 
In the following, 
in order to show more clearly the morphological and kinematical properties
of the simulated LMC,
we set the face-one view (edge-on view) of the LMC disk
to be always the $x$-$y$ ($x$-$z$) plane
by rotating the LMC disk by some degrees 
at  a given time $T$. 
Thus it should be noted in the following that the $x$-$y$ 
($x$-$z$ and $y$-$z$) plane is not identical with
the $X$-$Y$ ($X$-$Z$ and $Y$-$Z$, respectively) in the Figure 1.

\section{Results}

\subsection{The fiducial model}

\subsubsection{Formation of bar and thick disk}

Figures 5,  6, and 7  describe the morphological evolution of the LMC 
for the last $\sim$ 9 Gyr in
the fiducial model.
Owing to the smaller mass of the LMC ($M_{\rm LMC}$ = $10^{10}$ $M_{\odot}$),
the LMC disk is influenced by the strong tidal field of the Galaxy
from the early dynamical evolution of the LMC.
During the first pericenter passage of the LMC 
with respect to the Galaxy ($T$ $\sim$  $-6.8$ Gyr),
two spiral arms composed of old stars and gas
are formed within the disk owing to the tidal perturbation
from the Galaxy.
The distance between the Clouds is larger than $\sim$ 100 kpc
for $T$ $<$ $-6$ Gyr so that the SMC does not  dynamically
influence the LMC in the early dynamical evolution of the LMC.
After the first pericenter passage,
the bar-like structure composed mostly of old stars is formed in 
the central region of the disk ($T$ = $-5.5$ Gyr).  

Formation of field stars starts gradually from the central region
of the disk, in particular, from the high density regions of gas
along the inner spiral arms of the disk ($T$ =  $-7.1$ Gyr).
As the LMC first passes by the pericenter of its orbit with
respect to the Galaxy at $T$ = $-6.8$ Gyr,  the gaseous regions 
where field stars are actively forming
are  shifted from the center  to the two remarkable spirals
formed by the Galactic tidal perturbation.
After the formation of the central bar-like structure ($T$ = $-5.5$ Gyr),
the  gas density within the bar-like structure 
becomes higher owing to the
enhanced rates of cloud-cloud collisions along/within
the bar-like structure. 
As a result of this,
the field star formation becomes more efficient within
the central bar-like structure. 
The LMC disk is thus morphologically classified as
a barred spiral in the early phase of its evolution
($T$ $<$  $-5.5$ Gyr),
where the SMC does not dynamically influence  the LMC.

As the strong tidal interaction 
between the Clouds starts ($T$ = $-3.8$ Gyr),
the stellar bar grows and thus becomes more remarkable compared with
the outer stellar disk ($T$ = $-3.3$  Gyr). 
The formation of the stronger bar is associated closely with
(1) the tidal perturbation from the SMC 
and (2) formation of new stars along/within the bar. 
A significant fraction of old stars ($\sim$ 17 \%) 
initially within the outer part of the disk
are tidally stripped 
and field star formation is still ongoing mostly within the bar
after the first LMC-SMC encounter.
Consequently, most of old and new stars are located in the central bar
of the disk at $T$ = $-2.2$ Gyr.
Tidal interaction between the Clouds and the Galaxy
finally results in the formation of an elliptic disk 
which surrounds the central bar,
consists mainly of old stars,
and has a major axis misaligned significantly with
that of the central bar ($T$ = 0 Gyr).

\begin{figure}
\psfig{file=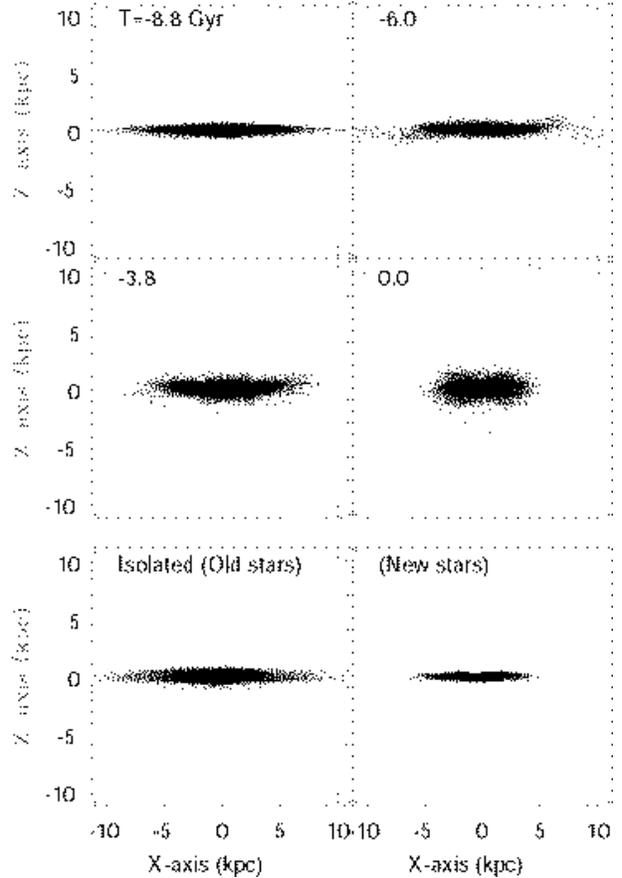,width=8.cm}
\caption{
The same as Figure 5 but seen from  the edge-on (the $X$-$Z$ plane)
for four different epochs.
For comparison, the final mass distributions
of old stars (left) and new stars (right) seen from 
the edge-on view in the isolated disk
model at $T$ = 0 Gyr are shown
in the bottom two panels.
Note that owing to the tidal stripping of outer halo stars by the
Galactic strong tidal field, the LMC slowly develops its stellar halo
with an inhomogeneous density distribution.
Note also that after the stellar bar formation ($T$ = $-5.5$ Gyr),
a very thick disk is finally formed.
The thick disk is significanly thicker than
the old stellar disk and the new stellar one
in the isolated model, which implies that tidal interaction between
the Clouds and the Galaxy is a main cause of the thick disk formation
(Compare the fiducial model
with the isolated one).
}
\label{Figure. 7}
\end{figure}

\begin{figure}
\psfig{file=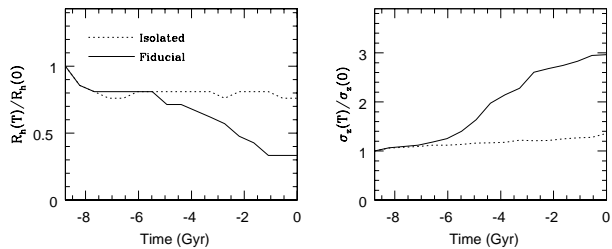,width=8.cm}
\caption{
The time evolution of the half-mass radius $R_{\rm h}$
and the vertical velocity dispersion (${\sigma}_{\rm z}$)
for the fiducial model (solid) and the isolated one (dotted).
The value of $R_{\rm h}$ (${\sigma}_{\rm z}$)
at each time step,  denoted as $R_{\rm h}(T)$ (${\sigma}_{\rm z}(T)$),
is normalized to the initial value  of $R_{\rm h}(0)$
(${\sigma}_{\rm z}(0)$)
at $T$ = $-8.8$ Gyr.
Note that $R_{\rm h}$ in the fiducial model
dramatically decreases owing to
the central new stars efficiently formed during the
tidal interaction between the Clouds and the Galaxy.
Note also that (1) ${\sigma}_{\rm z}$ also significantly
increases for $-6$ $\le$ $T$ $\le$ $-2$ Gyr in the fiducial
model
and (2) the degree of the increase is much more significant
in the fiducial model than in the isolated one.
These two results suggest that the increase of ${\sigma}_{\rm z}$
is not due to the numerical heating caused by small number
of stellar particles but due to the tidal heating from the Galaxy
and the SMC.
}
\label{Figure. 8}
\end{figure}

\begin{figure}
\psfig{file=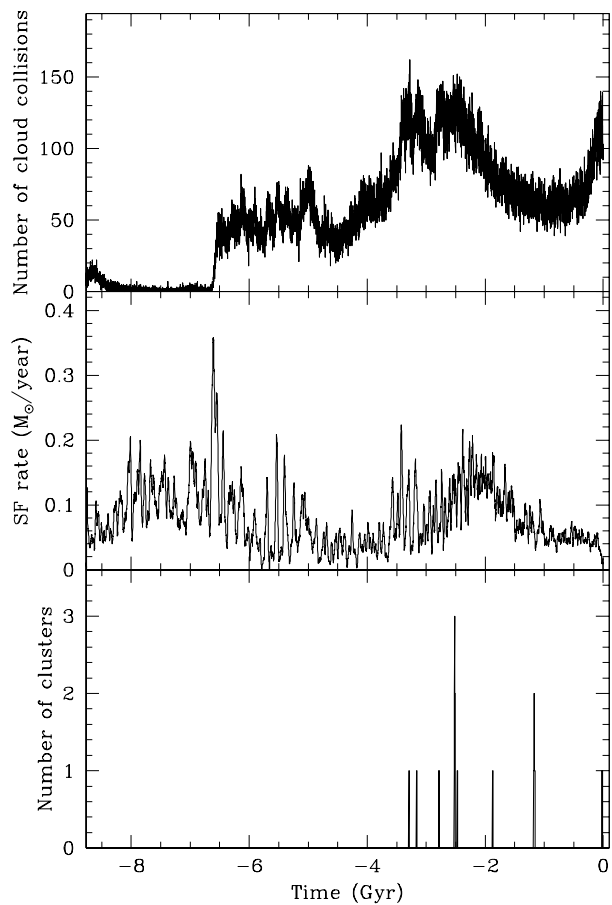,width=8.cm}
\caption{
Time evolution of cloud-cloud collision number (upper),
field star formation rate (middle),
and globular cluster formation rate (lower) in the fiducial model.
}
\label{Figure. 9}
\end{figure}

As is shown in Figure 7,
the initially thin stellar disk is finally transformed
into a thick one with the size significantly smaller than
its original one owing to the tidal interaction between
the Clouds and the Galaxy for the last $\sim$ 9 Gyr. 
The strong Galactic tidal field 
dynamically heats up the stellar disk every time
the LMC passes by the pericenter of its orbit with respect
to the Galaxy (e.g., $T$ $\sim$ $-6.8$ Gyr).
The vertical heating by the Galaxy 
and the resultant thickening of the  LMC disk 
starts from the outer, more fragile part of the disk.
After the stellar bar formation ($T$ = $-5.5$ Gyr),
the disk becomes severely warped and shows a sign of
a ``banana'' shape  at $T$ = $-3.8$ Gyr.
The bending instability following bar formation in tidal galaxy interaction
also contributes significantly to the thickening of a disk 
(e.g., Miwa \& Noguchi 1999).
The thickening of the LMC disk seen in Figure 7
for $-3.8$ $\le$ $T$ $\le$ 0 Gyr
is due partly to the bending instability in the disk.

As the disk thickening continues,
old stars stripped from the
outer part of the disk ($R_{\rm L}$ $>$ 5 kpc)
gradually develops  the stellar halo around the LMC disk. 
Although the distribution of stars in the halo
is quite elongated and inhomogeneous at $T$ = $-3.8$ Gyr,
it becomes more spherical and homogeneous at $T$  =  0 Gyr.
About 17 \% of the initial old stars are spatially redistributed
to form the outer stellar halo and only 2 \% of the halo
is composed of new stars formed from gas and then tidally stripped during the interaction
between the Clouds and the Galaxy.
One of the important characteristics of the stellar halo
is that the mass fraction of the halo to the disk
finally becomes an order of  $\sim$ 10 \%,
which is significantly higher than that of the Galaxy
(a few percent; e.g., Freeman 1987).   
The physical properties of  the stellar halo formed
from the Galaxy-LMC-SMC interaction will be discussed later.

Figure 8 summarizes the time evolution of global structural
and kinematical properties of the LMC stellar disk for
the last $\sim$ 9 Gyr. 
Firstly,  the effective radius represented by $R_{\rm h}$
(the half-mass radius) of the disk becomes significantly
small owing to the formation of new stars in the central
region of the disk. For example,
$R_{\rm h}$ shows $\sim$ 65 \% of the original
value for $T$ = $-3.8$ Gyr (before the commencement of
the strong tidal interaction with the SMC)
and finally shows $\sim$ 33 \% of the original one
at $T$ = 0 Gyr: The disk size becomes smaller as the disk
is dominated more significantly by young stellar populations.
The isolated model shows only $\sim$ 22 \% decrease of its
disk size owing to the much less efficient star formation
in its central region, which confirms that the dramatic
change of the size of the LMC disk in the fiducial model is caused by 
central star formation triggered by the Galaxy-LMC-SMC interaction.

As is shown in Figure 8, the vertical velocity dispersion represented by
${\sigma}_{\rm z}$ dramatically (a factor of $\sim$ 3) increases during
the Galaxy-LMC-SMC interaction  for the last $\sim$ 9 Gyr, whereas
${\sigma}_{\rm z}$ of the disk in the isolated disk model
only gradually increases by only a factor of $\sim$ 1.4.
The increase in ${\sigma}_{\rm z}$ for the isolated model
is caused by (1) the scattering of stars via gas clouds,
(2) ``self-heating'' by the spiral arms in the disk
(e.g., Sellwood \& Carlberg 1984), 
and (3) numerical heating due to the adopted
small particle number (an order of $10^5$) in the present simulation.
Given the fact that the increase in ${\sigma}_{\rm z}$
in the fiducial model is more than a factor of 2 larger 
than that in the isolated one,
the contribution of numerical heating to ${\sigma}_{\rm z}$
in the fiducial model is quite minor.
Therefore, it is reasonable to say that 
the increase in ${\sigma}_{\rm z}$ in the fiducial model is caused mainly by
the tidal heating from the Galaxy and the SMC.

\begin{figure}
\psfig{file=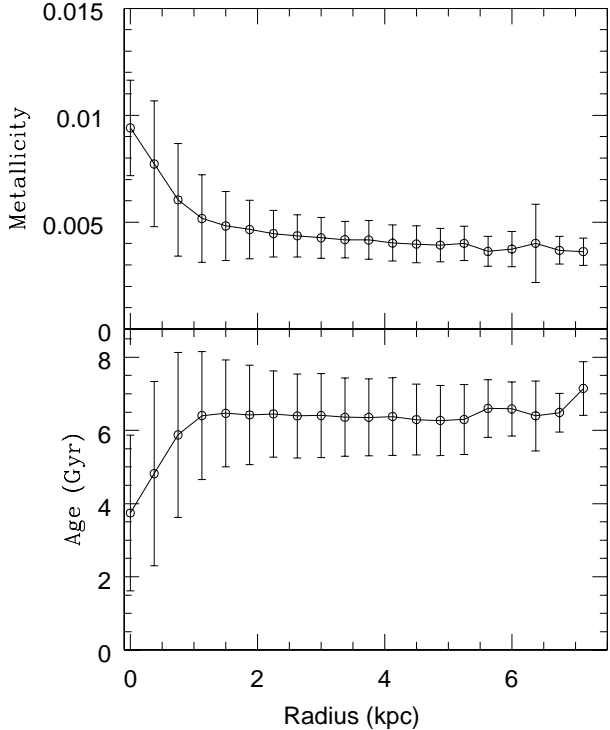,width=8.cm}
\caption{
The radial gradients of age (lower) and metallicity (upper)
in the LMC disk at $T$ = 0 Gyr in the fiducial model.
}
\label{Figure. 10}
\end{figure}

\begin{figure}
\psfig{file=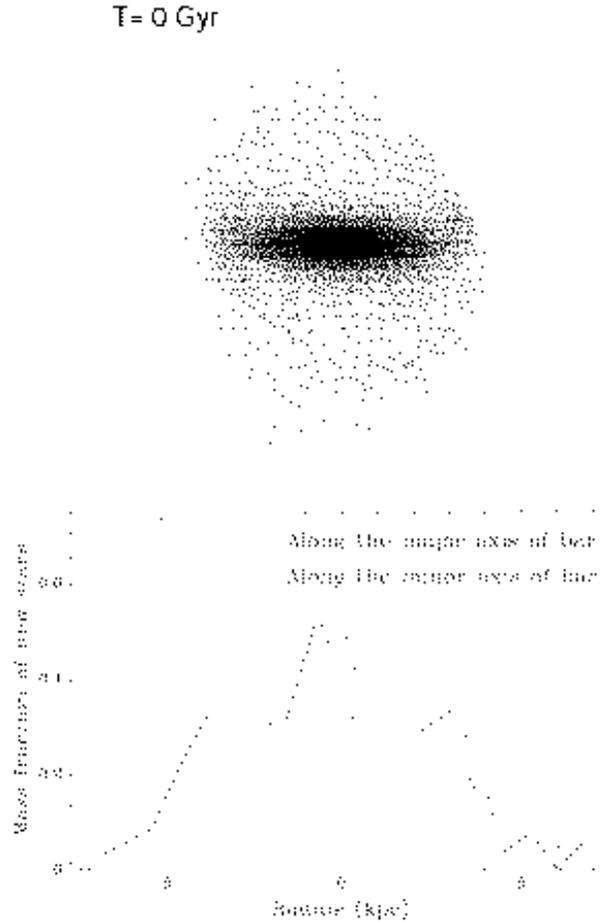,width=8.cm}
\caption{
Distribution  of all particles seen from the face-on view of the LMC disk
(upper) and the radial distribution of the mass fraction of new stars
to all stars (lower)
at $T$ = 0 Gyr in the fiducial model.
The mass fraction of new stars is estimated along the major axis
of the stellar bar (solid) and along the minor axis (dotted).
}
\label{Figure. 11}
\end{figure}

\subsubsection{Stellar populations}

Figure 9 shows the formation histories of field stars and GCs
in the fiducial model. 
The formation rate of field stars gradually increases
to reach $\sim$ 0.1 $M_{\odot}$ yr$^{-1}$ in
the first $\sim$ 1 Gyr dynamical evolution owing to the 
development of high density gas regions along the spiral arms
in the disk. 
After the first pericenter passage of the LMC ($T$ $\sim$ $-6.8$ Gyr),
the star formation rate is moderately enhanced 
and reaches  $\sim$ 0.4 $M_{\odot}$ yr$^{-1}$
(averaged for 0.1 dynamical time of the LMC  corresponding to $\sim$ 13 Myr) 
owing to the 
formation of strong, double-armed gaseous arms, where
gas density becomes significantly high. 
The cloud-cloud collision rate is dramatically (more than
an order of magnitude) enhanced  within $\sim$ 0.5 Gyr    
after the pericenter passage, because the strong Galactic
tidal effects increase the velocity dispersion of the gas clouds. 
However, the cloud-cloud collisions that
leads to the formation of GCs in the disk do not occur
in this first pericenter passage, 
just  because the tidal perturbation is not strong enough
to trigger cloud-cloud collisions with moderately high relative velocities
(between 30 and 100 km s$^{-1}$) and small impact parameters ($<$ 0.25).

Owing to the rapid gas consumption by formation of field
stars during/after the first pericenter passage of the LMC,
the formation rate of the field stars does not increase significantly
until the strong Galaxy-LMC-SMC tidal interaction 
begins at $T$ $\sim$ $-3.8$ Gyr.
The tidal perturbation from the SMC and  the Galaxy
triggers the moderately enhanced star formation rate 
of $\sim$ 0.1 $M_{\odot}$ for  $-3.5$ $<$ $T$ $<$ $-2$  Gyr.
During this period,  the GC formation also becomes efficient,
essentially because the combined tidal effects of the Galaxy
and the SMC are strong enough to enhance cloud-cloud
collisions required for GC formation in the LMC disk.
The peak of the GC formation ($\sim$ $T$ $\sim$ $-2.5$ Gyr)
is nearly coincident with that of the field stars for 
$-3.5$ $<$ $T$ $<$ $-2$ Gyr.
The GC formation also occurs at $T$ $\sim$  $-0.2$ Gyr,
when the LMC-SMC distance becomes very small (less than 10 kpc;
smaller than the original LMC disk size) so that the LMC
collides with the SMC.
This final collision between the Clouds around $0.2$ Gyr ago
could have significant effects in the recent star formation
histories of the LMC, as suggested by previous authors 
(e.g., GSF). 

About 47 \% of the initial gas is converted into new field stars
for the last $\sim$ 9 Gyr evolution of the LMC and most of new
stars are concentrated in the central bar.
Only 0.5 \% of the gas is converted into GCs, which reflects
the fact that the cloud-cloud
collisions with moderately high speed (between 30 and 100 km s$^{-1}$)
and small impact parameter ($<$ 0.25), required for cluster formation,
do not occur until the LMC begins to interact violently with the
SMC when the two are less than 10\,kpc apart (T=$-3.6$\,Gyr). 
Chemical enrichment resulting from  the moderately enhanced star formation 
increases gradually the metallicity of new field stars and GCs.
For the fiducial model with the initial gaseous metallicity of 0.002
([Fe/H] = $-1.0$), $y_{\rm met}$ (chemical yield) of 0.005, 
and $R_{\rm met}$ of 0.3, the  mean metallicity of
new field stars within the disk and the halo 
finally becomes 0.007 ($-0.46$ in [Fe/H]) 
at $T$ = 0 Gyr (It should be emphasized here that the
final metallicity depends  strongly on the 
initial values of the initial metallicity, $y_{\rm met}$,
and $R_{\rm met}$).

\begin{figure}
\psfig{file=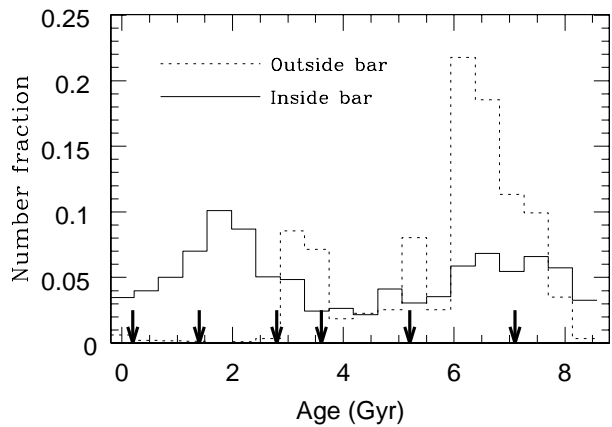,width=8.cm}
\caption{
The age distribution of new stars inside the stellar bar (solid)
and outside the bar (dotted) at $T$ = 0 Gyr in the fiducial model.
A new star is  regarded as  ``inside the bar'',
if the distance of the star along the major axis of the bar
is less than 5 kpc {\it and} if that along the minor axis
are is less than 1.5 kpc.
For convenience, the normalized number of new stars is
shown.  The epochs of the pericenter passage of the SMC with
respect to the LMC are shown by thick arrows for comparison.
Note that the age distribution of new stars inside the bar shows
the peak around $\sim$ 2 Gyr whereas that outside the bar
shows the peak around 6 Gyr.
}
\label{Figure. 12}
\end{figure}

Chemical enrichment proceeds more
in the central region  where
gas consumption by star formation is more rapid and efficient 
owing to the gas inflow 
triggered by the Galaxy-LMC-SMC interaction.
The disk consequently shows a negative metallicity gradient
for new field stars  in 
the sense that the inner regions show higher metallicity
than the outer ones at $T$ = 0 Gyr (Figure 10).
The disk also shows a positive age gradient 
with the new field stars within the central 1 kpc
of the disk being  a few Gyr  younger than
those outside the central 1 kpc. 
Both  age and metallicity gradients are rather flat 
for 3 $\le$  $R_{\rm L}$ $\le$ 7 kpc in the disk
(where $R_{\rm L}$ is the projected distance from
the center of the LMC disk),
possibly  because the stellar bar
dynamically mixes the young stellar population
with different ages and metallicities owing to the stream
motion of stars and gas  along the bar.
Both the age and metallicity dispersions
are larger in the central regions, which reflect the fact
that gaseous components with different metallicities
are transferred to the center and converted to new stars
there in different epochs during the Galaxy-LMC-SMC
interaction. It should be stressed here that
recent observations have found 
that intermediate-age stars are more centrally concentrated
than the older stars in the LMC disk (e.g., Cole et al. 2000).

\begin{figure}
\psfig{file=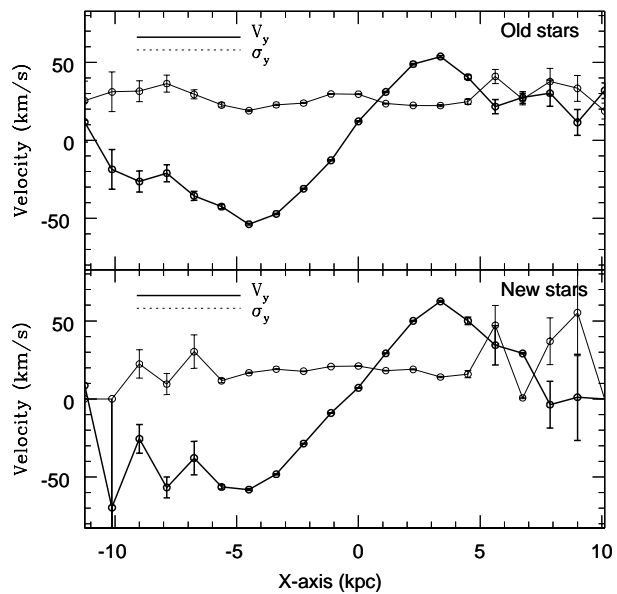,width=8.cm}
\caption{
Radial dependences of line-of-sight-velocity ($V_{\rm y}$)
and velocity dispersion (${\sigma}_{\rm y}$) along with
the $x$-axis  for old stars (upper)
and for new stars (lower) at $T$ = 0 Gyr in the fiducial model.
}
\label{Figure. 13}
\end{figure}

\begin{figure}
\psfig{file=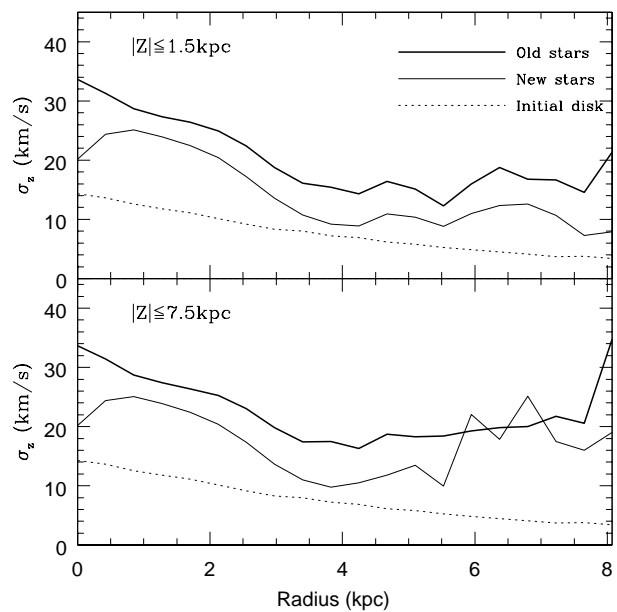,width=8.cm}
\caption{
Radial dependences of vertical velocity dispersion (${\sigma}_{\rm z}$)
for old stars (thick solid) and new stars (thin solid)
with the vertical distance $|z|$ from the disk plane of the LMC
less than 1 kpc (upper) and those with $|z|$ less than 7.5 kpc (lower)
at $T$ = 0 Gyr in the fiducial model.
For comparison, the results of the isolated model are also shown
by dotted lines.
}
\label{Figure. 14}
\end{figure}

\begin{figure}
\psfig{file=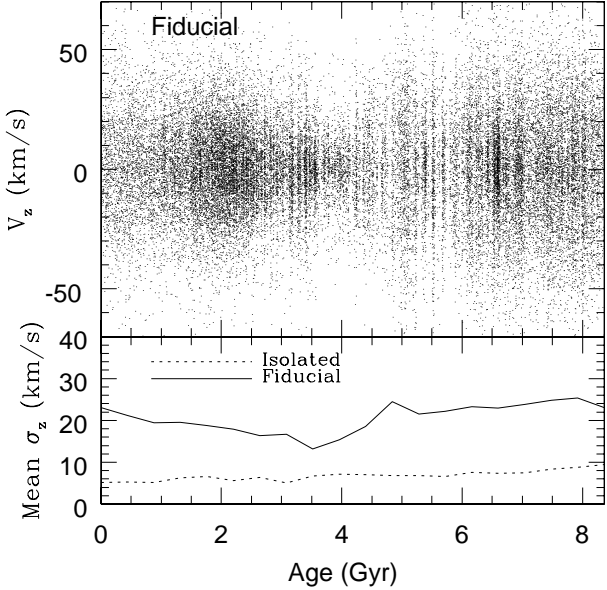,width=8.cm}
\caption{
The distribution of new stars on the age-$V_{\rm z}$ plane,
where $V_{\rm z}$ is the line-of-sight velocity parallel to  the $z$-axis
(upper) and the mean value of
the vertical velocity dispersion ${\sigma}_{\rm z}$
as a function of age for the new stars  (lower)
at $T$ = 0 Gyr in the fiducial model.
The lower figure therefore describes the age-dispersion
relation of new stars. For comparison, the age-dispersion
relation for the isolated model is shown by a dotted line
in the lower panel.
}
\label{Figure. 15}
\end{figure}

Figure 11 clearly shows that the mass fraction of new field stars
(referred to as $f_{ns}$ hereafter) along the major axis
of the bar 
is significantly larger within the stellar bar than outside it.
About 50 \% of the stars within the central $\sim$ 500 pc of the disk
along the major axis of the bar
are new field stars formed from gas during the Galaxy-LMC-SMC interaction.  
The radial profile of $f_{ns}$
shows a relatively flat density profile (``shoulder'') with
$f_{\rm ns}$ $\sim$ 0.3 (0.2)
along the major (minor)  axis of the bar
and has abrupt steepening of the gradients around $|R_{\rm L}|$ = 1 kpc 
and 4 kpc.
The radial gradient for $|R_{\rm L}|$  $\le$  7 kpc
appears to be slightly steeper
along the minor axis of the bar than along the major one.
$f_{\rm ns}$ becomes  smaller than 0.1 for 5 kpc $\le$ $|R_{\rm L}|$,
which indicates that the outer disk is dominated by old stellar populations.
These radial dependences of $f_{\rm ns}$
are due essentially to the centrally concentrated young
stellar populations formed during the Galaxy-LMC-SMC interaction.

Figure 12 shows that the stellar age distribution 
is significantly different between new stars  inside
the bar and those outside the bar.
The stellar age distribution inside the bar has a peak 
around $\sim$ 2 Gyr, which means that the stellar population 
inside the bar is dominated by new stars that are formed {\it after}
the  Galaxy-LMC-SMC interaction becomes stronger owing
to the dynamical coupling of the Clouds.
The bump around the ages of $6-8$ Gyr in the age distribution
inside the bar suggests that 
some fraction of new stars formed in the early Galaxy-LMC
interaction.
The age distribution outside the bar, on the other hand,
has a peak around 6 Gyr, which means that the stellar
population outside the bar is dominated by new stars
formed in the early evolution of the LMC, in particular,
those formed during the Galaxy-LMC interaction $6-7$ Gyr ago. 
These results in Figures 11 and 12 
suggest that (1) the dominant stellar population
is significantly  different between different regions of the LMC disk
and (2) the difference is due essentially  to
the Galaxy-LMC-SMC interaction which forms a bar   
and thus drives efficient inner transfer of interstellar gas.

\subsubsection{Kinematics}

The long-term tidal perturbation from  the Galaxy and the SMC to the LMC
causes dramatic changes in kinematical properties of the disk
that is initially ``dynamically cold''.
Figure 13 shows that the rotational velocity  of old stars
(hereafter referred to as $V_{\rm rot}$, and represented as $V_{\rm y}$
in Figure 13 for convenience) has a peak value of $\sim$ 50 km s$^{-1}$ 
around the central $3-4$ kpc and 
then decreases sharply toward outward from there.
The ratio of the maximum value of $V_{\rm rot}$ 
to $V_{\rm m}$ (the maximum value of
the circular velocity $V_{\rm c}$ in the initial disk)
for old stars is $\sim$  0.7, 
which implies that azimuthal/radial  velocity dispersion  of old stars
is significantly increased by the Galaxy-LMC-SMC interaction. 
$V_{\rm rot}$ decreases outside the central $3-4$ kpc
much more sharply than expected from the initial mass profile
of the LMC. The initial disk shows a decrease
by a factor of  $\sim$ 17 \%  in $V_{\rm c}$ (and thus $V_{\rm rot}$) 
for 5 $\le$ $R_{\rm L}$ $\le$ 10 kpc 
whereas $V_{\rm rot}$ in the final LMC disk shows
a decrease by a factor of  $\sim$ 60 \% in $V_{\rm rot}$
for the corresponding region. 
This clearly indicates that the long-term dynamical heating of the LMC disk
by the Galaxy and the SMC changes the shape of the rotation curve
of the LMC.

The initial radial profile of azimuthal velocity dispersion  of old stars
is monotonously decreasing toward the outer region of the LMC disk.
As shown in Figure 13, the velocity dispersion ${\sigma}_{\rm y}$
(a measure of azimuthal velocity dispersion) shows a large value
of $30-40$ km s$^{-1}$,
which is even larger than the central one of $\sim$ 20 km s$^{-1}$,
around $7-8$ kpc from the center of the disk. 
This result confirms that the long-term dynamical heating 
by the Galaxy and the SMC drives the kinematical change
of the LMC's outer disk that is more susceptible to 
external tidal perturbation.
The results of the new stars are essentially the same as those
of old ones, except that (1) the velocity dispersion ${\sigma}_{\rm y}$
is {\it on average}  smaller in new stars than in old ones
and (2) the maximum  $V_{\rm rot}$ is only slightly
larger in new stars than in old stars probably because of 
gaseous dissipation.
The asymmetric profiles of ${\sigma}_{\rm y}$ and $V_{\rm rot}$
seen both in old stars and in new ones
are due partly to the collision between the Clouds at
$T$ $\sim$ $-0.2$ Gyr.

Figure 14 shows that the radial gradient of the vertical velocity
dispersion ${\sigma}_{\rm z}$ of the disk for stars with
the vertical distance ($|z|$) from  the disk plane equal to or less than
1.5 kpc (i.e., those stars within the thick disk). 
It is clear from this Figure 14 that 
${\sigma}_{\rm z}$  of old stars shows a monotonous decrease
for $R_{\rm L}$ $\le$ 4 kpc and begins to increase 
gradually and slightly toward outward from there, 
though the profile for $R_{\rm L}$ $>$ 4 kpc is somehow irregular.
The outwardly increasing ${\sigma}_{\rm z}$ is more remarkable
for old stars with $|z|$ $\le$ 7.5 kpc,
because these stars  include  some fraction
of old halo stars that form a  kinematically hot stellar system
around the LMC.
The larger ${\sigma}_{\rm z}$ in the outer LMC disk means that
the scale height also increases with radius for $R_{\rm L}$ $>$ 4 kpc. 
These results are broadly consistent with recent observational results
by Alves \& Nelson (2000) and vdMAHS.

${\sigma}_{\rm z}$ is increased by a factor of 2.4 in the center
of the disk and a factor of 5.5 at $R_{\rm L}$ =  8 kpc
for the old stars with $|z|$ $\le$ 1.5 kpc
owing to  the long-term tidal interaction  between the Galaxy and the Clouds.
The radial profiles of ${\sigma}_{\rm z}$ for new stars also
show such an outwardly increasing ${\sigma}_{\rm z}$ 
for $R_{\rm L}$ $>$ 4  kpc, though the increase is less remarkable
compared with that seen in old stars.
${\sigma}_{\rm z}$  of new stars is on average lower than
that of old stars for nearly every radii,
essentially because new stars are formed from gas that
dissipates away random kinematical energy yielded  by
the Galaxy-LMC-SMC interaction. 
The central ``dip'' in the radial profile of ${\sigma}_{\rm z}$
for new stars is also caused by efficient gaseous dissipation there.
Thus Figures 13 and 14 suggest that {\it the outer kinematics
of the LMC disk have valuable information on
the past interaction history of the LMC with the Galaxy and the SMC}.

\begin{figure}
\psfig{file=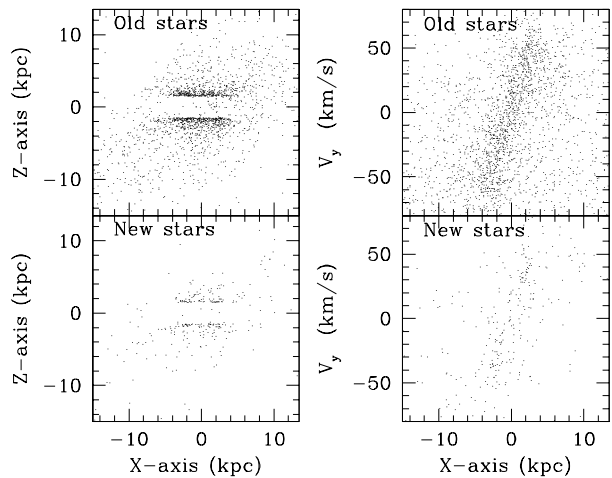,width=8.cm}
\caption{
The distribution of
halo old stars (upper left) and halo new ones (lower left)
seen from the edge-on of the LMC disk
and the radial dependences of the line-of-sight
velocity parallel to the $y$-axis ($V_{\rm y}$ for
halo old stars (upper right) and for halo new stars (lower right)
at $T$ = 0 Gyr in the fiducial model.
Here a star can be classified as a ``halo'' star
either if
the projected distance from the center of the LMC disk
is less than 7.5 kpc
or if the vertical distance from the LMC disk plane
is less than 1.5 kpc.
}
\label{Figure. 16}
\end{figure}

\begin{figure}
\psfig{file=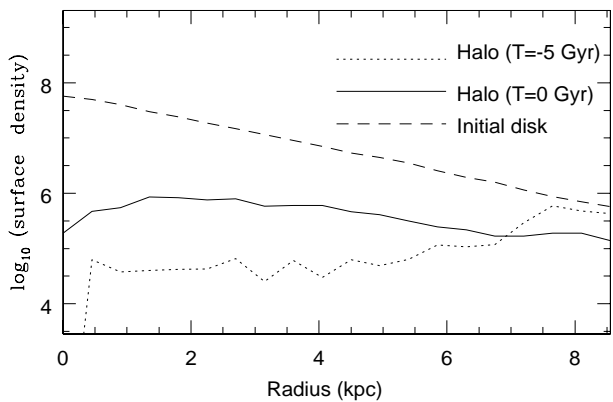,width=8.cm}
\caption{
The  projected density distribution of halo stars
at $T$ = 0 Gyr (solid) and at $T$ = $-5$ Gyr (dotted)
in the fiducial model.
For comparison, the radial distribution
of stars  in the initial exponential disk is shown by
a dashed line for comparison.
The definition of ``halo stars'' is the same as that described
in Figure 16.
}
\label{Figure. 17}
\end{figure}

Figure 15 describes a  relation between velocity dispersion
of ${\sigma}_{\rm z}$ and ages (i.e., so-called ``age-dispersion relation'')
for new stars of the LMC disk.  
The isolated model shows the trend of decreasing ${\sigma}_{\rm z}$
with decreasing ages, which reflects the fact that a new star
formed more recently originates from gas that has experienced  a larger amount
of gaseous dissipation. 
This tendency is seen in the disk of the fiducial model
for the new stars with the ages older than $\sim$ 4Gyr,
though the mean ${\sigma}_{\rm z}$ is more than a factor of 2 
higher in the fiducial model than in the isolated model
owing to the dynamical heating by the Galaxy.
However, the new stars with the ages younger than $\sim$ 4 Gyr
show an interesting tendency of ${\sigma}_{\rm z}$ slightly
increasing with decreasing ages.
This is probably because a new star formed more recently 
originates from gas that is more strongly randomized by
{\it the combined tidal effects of the Galaxy and the SMC}:
The net effect of gaseous dissipation is  weaker than 
that of  randomization of gaseous motion by the stronger
tidal perturbation from the Galaxy and the SMC
for the last $\sim$ 4 Gyr.
These results imply that the age-dispersion relation
in stellar populations of the LMC disk
is significantly different from that observed in
the Galaxy.

\begin{figure}
\psfig{file=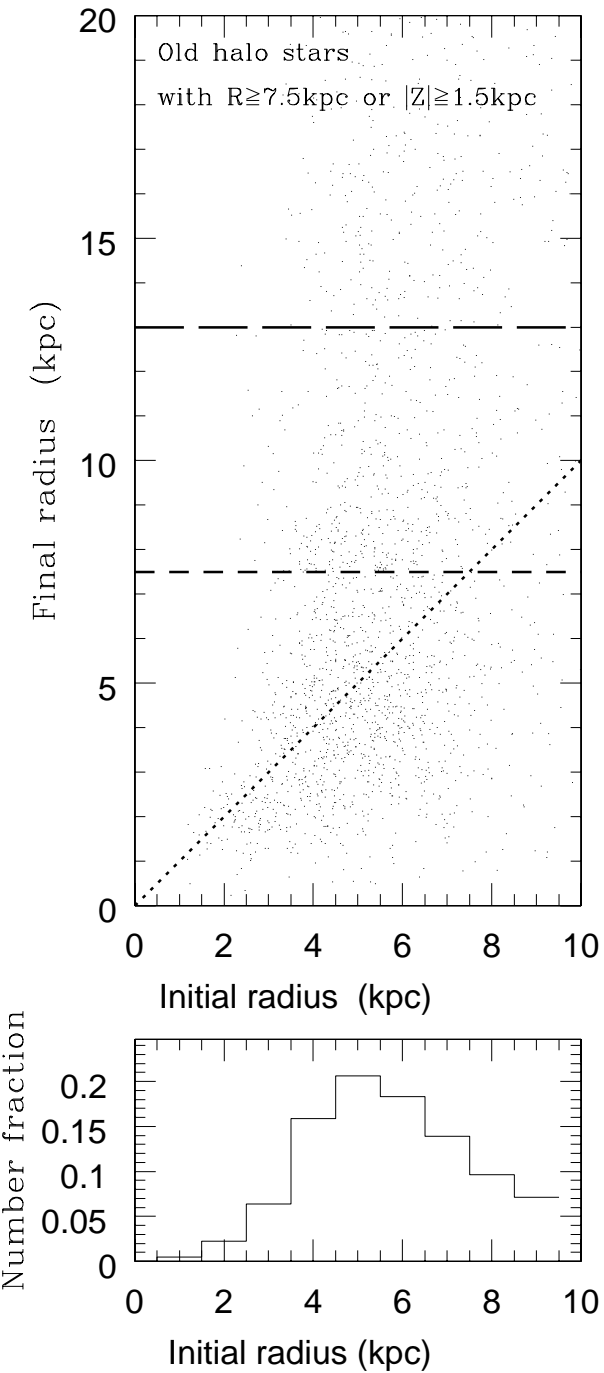,width=8.cm}
\caption{
Final radii of halo old stars as a function of the initial radii
(upper) and the number distribution of the initial radii of the stars
that are regarded as ``halo stars''
at $T$ = 0 Gyr (lower) in the fiducial model.
The definition of ``halo stars'' here is the same as described
in Figure 16.
The long-dashed and short-dashed lines represent
the initial tidal radius $r_{\rm t}$ (= 13 kpc)
and $\sim$ 3$R_0$ (7.5 kpc),
where $R_0$ is the scale radius of the initial LMC disk,
respectively.
Stars above the dotted line are those for which the final radii
are larger than the initial ones (i.e., they are transfered outward
due to the tidal effects of the Galaxy and the SMC).
For convenience, the normalized number (i.e. number fraction)
is shown in the lower panel.
Note that most of the halo stars originate from the outer
part ($R$ $>$ 4 kpc) of the initial LMC disk.
}
\label{Figure. 18}
\end{figure}

\subsubsection{Stellar halo properties}

One of the important outcome of the long-term
tidal interaction between the Clouds and the Galaxy
is the formation of a stellar halo through the redistribution
of old stars initially within the LMC disk.
It should also be stressed here that our disk models 
initially do not include very old ($\sim$ $10-13$ Gyr),
metal-poor ([Fe/H] $<$ $-1.6$)
stellar halo components that the Galaxy is observed to have.
Therefore, the halo properties in the present
study are dependent totally
on those of old stars and new stars stripped from
the original disk.
Accordingly, direct comparison of the present numerical
results with observations is not simply possible,
because the LMC stellar halo may contain the
old stellar halo that was formed {\it prior to} its disk formation.
We thus suggest that the  present results may well be
able to be compared with observational results
for halo stars with ages less than
$\sim$ 9 Gyr.

As have been shown in Figure 7,
some fraction of old stars stripped from the outer part of
the disk due to the Galaxy-LMC-SMC interaction
are redistributed in the outer halo region
of the disk. 
Figure 16 shows structural and kinematical properties of
the stellar halo composed of stars either outside
the disk (i.e., $R_{\rm L}$ $>$ 7.5 kpc) or well above/below
the thick disk (i.e., $|z|$ $>$ 1.5 kpc).
Although these old halo stars have a relatively homogeneous
distribution within the central $\sim$ 6 kpc,
they have an elongated  distribution 
for a wider field  (12 kpc $\times$ 12 kpc) of view
owing to the old stars being now dynamically influenced by
the tidal fields  of the Galaxy and the SMC. 
The new stars in the halo also show such an elongated
distribution, though it is less remarkable compared with
old halo stars. 
The new halo stars appear to be more flattened 
compared with old halo ones for the central $\sim$ 6 kpc
because of the smaller number of the stars with $|z|$ $>$ 4.0 kpc.
Since the mass fraction of the new stars
in the halo region is only $\sim$ 2 \%,
the entire  distribution of  the halo 
is determined by old halo stars: The halo has an inner 
homogeneous distribution and an outer elongated 
(thus inhomogeneous) one.

As shown in Figure 16,
both old and new halo stars show a sign of rotation,
in particular, for those within the central 2 kpc,
if they are seen from the edge-on view. 
Although the inner stellar components  with a certain
amount of rotation should be regarded as parts of
the thick disk rather than the halo,
the moderately rotating inner halo
is regarded as an important  characteristic that the stellar halo
has if it originates from the outer part of the initially thin stellar disk
through the redistribution of the stars during 
the Galaxy-LMC-SMC interaction. 
Figure 17 shows that (1) the projected density of the stellar halo
at $T$ = 0 Gyr is roughly approximated as an exponential
profile with the slope shallower than the original exponential disk
for $R_{\rm L}$ $>$ 2 kpc,
(2) the projected density of the stellar halo 
is on average more than
an order of magnitude lower than the original stellar disk
and the difference in the density depends strongly on the radius,
and (3) the projected density profile changes significantly with
time at a given  radius and it becomes steeper as the time passes by.

Figure 18 demonstrates that about 75 \% of old stars in 
the halo (at $T$ = 0 Gyr) originate from the disk
regions with the distances larger than 5 kpc.
This is because the tidal stripping of old disk stars
and the subsequent redistribution of the stripped stars
more efficiently occur during the Galaxy-LMC-SMC
interaction. 
As a natural result of this,  the final metallicity
distribution function (MDF) in the old halo stars is
expected to be significantly different from that of the initial disk.
Figure 19 shows a possible MDF of the old halo stars
for the fiducial model with an initial negative metallicity
gradient (described in the equation (11) in \S 2) and  
the mean metallicity of [Fe/H] = $-1$. 
The MDF in this Figure 19 clearly shows a peak around
[Fe/H] =  $-1.2$ $\sim$ $-1.3$, which is by $\sim$ $-0.3$  (in dex)
lower than that of the initial disk and is
by $\sim$ $-0.4$  (in dex)  higher than that of the observed
Galactic stellar halo (e.g., Freeman 1987).
The derived lower metallicity of the old halo stars is due
to the stars originating preferentially from
the outer disk regions where the stellar metallicity
is lower owing to the negative metallicity gradient.

The derived difference in the MDF between the simulated stellar halo  and 
the observed Galactic stellar halo probably reflects the fact
that the formation history of the Galactic stellar
halo is significantly different from what is described in
this paper (e.g., Bekki \& Chiba 2000, 2001).
One of the significant differences between the simulated
stellar halo and the observed Galactic one
is that the simulated halo includes a fraction of relatively metal-rich
($-1$ $<$ [Fe/H] $<$ $-0.3$) and moderately young stars, 
though the mass fraction  of these populations among the entire
halo population is very small ($\sim$ 2 \%).
These metal-rich halo components result from the tidal stripping
of new stars formed the LMC disk in the later phase of the
Galaxy-LMC-SMC interaction. Therefore, the detection of such
metal-rich, young stars within  the LMC halo region in
future observations could be an evidence that 
the Galaxy-LMC-SMC interaction is partly responsible for,
at least, some part of the LMC stellar halo.

\begin{figure}
\psfig{file=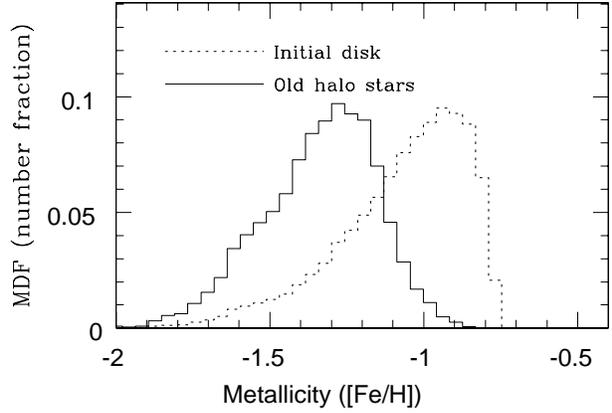,width=8.cm}
\caption{
The metallicity distribution function (MDF) of
halo old stars is shown by a solid line
for  $T$ = 0 Gyr in the fiducial model.
For comparison, the initial MDF of the LMC disk
is shown by a dotted line.
}
\label{Figure. 19}
\end{figure}

\begin{figure}
\psfig{file=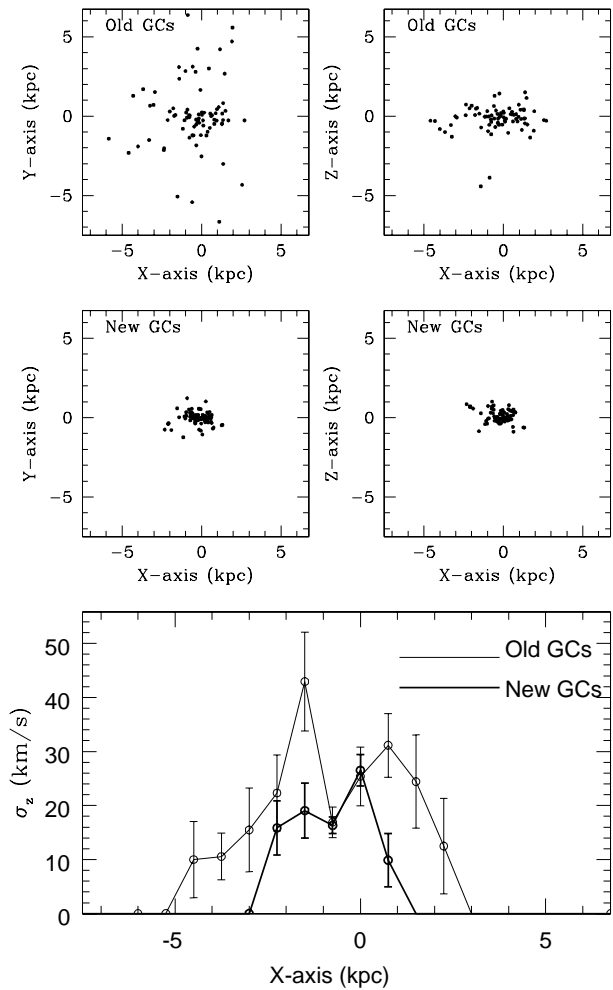,width=8.cm}
\caption{
The distribution of old GCs projected onto the $x$-$y$ plane
(top left)  and  onto the $x$-$z$ one (top right)
and that of new GCs
projected onto the $x$-$y$ plane (middle left)
and  onto the $x$-$z$ one (middle  right)
at  $T$ = 0 Gyr in the fiducial model.
The bottom panel shows the radial profiles
of vertical velocity dispersion (${\sigma}_{\rm z}$)
for old GCs (thin solid) and for new GCs (thick solid).
}
\label{Figure. 20}
\end{figure}

\begin{figure}
\psfig{file=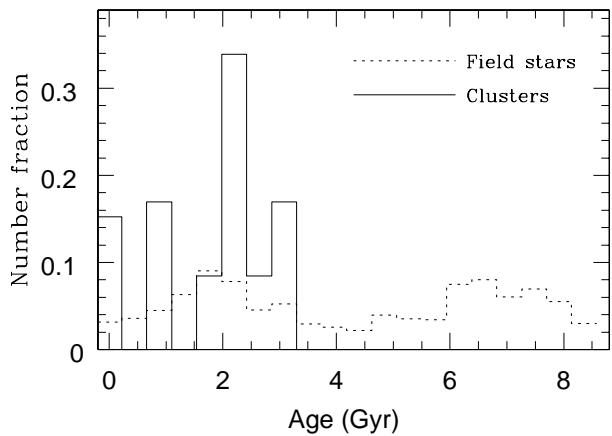,width=8.cm}
\caption{
Age distributions of field stars (dotted) and clusters
(solid) at  $T$ = 0 Gyr in the fiducial model.
For convenience, the normalized fraction of stars
in each age bin is shown.
}
\label{Figure. 21}
\end{figure}

\subsubsection{Properties of GCs}

Figure 20 summarizes structural and kinematical properties 
of old and new GCs in the fiducial model at $T$ = 0 Gyr. 
As shown in this Figure 20,
the final spatial distribution is remarkably different between
the old GCs and the new ones.
The new GCs has a more compact distribution than
the old GCs and most of them are concentrated  within the central
$\sim$ 3 kpc. This is essentially because the new GCs
form only in the high-density gaseous regions where 
the cloud-cloud collisions with moderately high speed 
(between 30 and 100 km s$^{-1}$)
and small impact parameter ($<$ 0.25), required for cluster formation,
occur.
The face-on distribution of the new GCs 
is fairly elongated with the direction of the elongation
nearly parallel to the major axis of the stellar bar:
Most of the new GCs are within the stellar bar in the disk.
Nearly all of the new GCs is confined within 1 kpc from
the LMC disk plane, which reflects the fact that 
the new GCs originate from the high-density regions in the thin gaseous disk.

The old GCs, on the other hand, show a more wide spread distribution
and are not necessarily confined in the central bar region of the disk.
The edge-on distribution of the old GCs shows a thick-disk appearance
with the two GCs having $|z|$ of $>$ 3 kpc and being able to be regarded
as halo GCs.  These two halo GCs are initially in the outer stellar
disk and thus tidally stripped and subsequently distributed in
the halo region. 
Figure 20 also shows a few differences in the final radial profile
of vertical velocity dispersion (${\sigma}_{\rm z}$)
between the two different GC populations,
though the final profiles of ${\sigma}_{\rm z}$
are severely influenced by the tidal perturbation
from the Galaxy and the SMC, in particular,
from the collision with the SMC around $T$ $\sim$ $-0.2$ Gyr
(Here the center of mass of {\it field stars} is adopted for
the estimation of kinematics).
${\sigma}_{\rm z}$ is systematically  higher in the old GCs
than in the new GCs throughout the disk,
because the old GCs experienced for a much longer time ($\sim$ 9 Gyr)
the tidal heating from the Galaxy (and the SMC)
than the new GCs all of which are formed relatively recently
($T$ $<$ $-3.3$ Gyr). 

Figure 21 describes the difference in the age distributions 
between the new field stars and the new GCs in the disk.
It is clear from this figure that
all  clusters have ages younger than $\sim$ 3.3\,Gyr
whereas the field star population show a wide distribution
of ages.  This result reflects the fact that
the field star formation is sensitive to local gas density whereas
the cluster formation occurs only when random motion
of gas in the LMC becomes significantly large.
The mean metallicity of the new GCs at $T$ = 0 Gyr 
is by $-0.08$ dex (in [Fe/H])
higher than that of the new field stars, 
because chemical enrichment associated 
with field star formation proceeds efficiently {\it prior to 
the formation of new GCs} (i.e., $T$ $<$ $\sim$ $-3$ Gyr)
and consequently the new GCs are thus formed from more metal-rich gas. 
The derived difference suggests that
if future observations on the detailed age distribution of field
stars for the entire disk region of the LMC
reveal the differences in age and metallicity distributions
between field stars and GCs,
the difference can be understood in terms of the difference
in the formation processes between field stars and GCs
during the Galaxy-LMC-SMC interaction.

\begin{figure}
\psfig{file=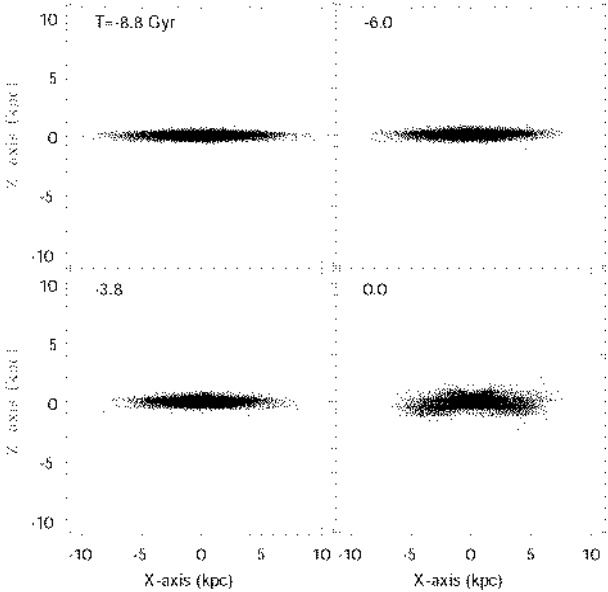,width=8.cm}
\caption{
The same as Figure 7 but for the model 9.
}
\label{Figure. 22}
\end{figure}

\begin{figure}
\psfig{file=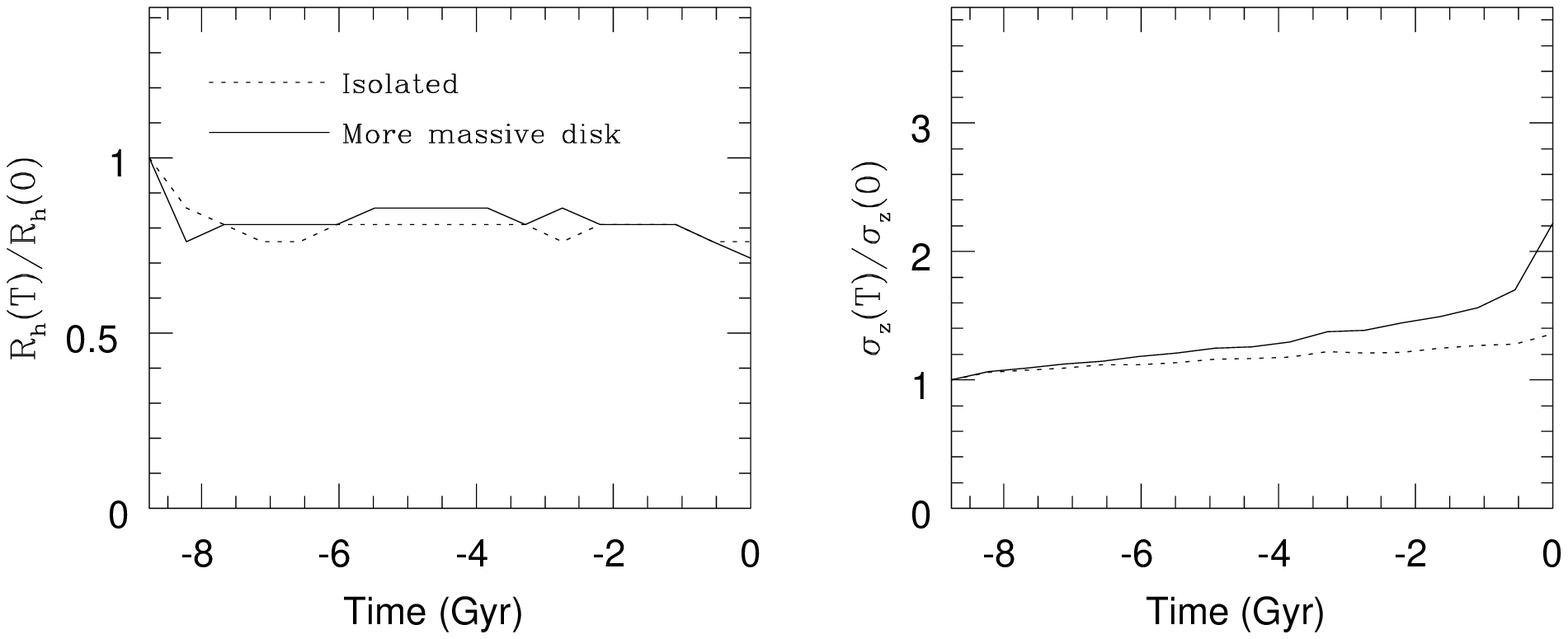,width=8.cm}
\caption{
The same as Figure 8 but for the model 9.
}
\label{Figure. 23}
\end{figure}

\begin{figure}
\psfig{file=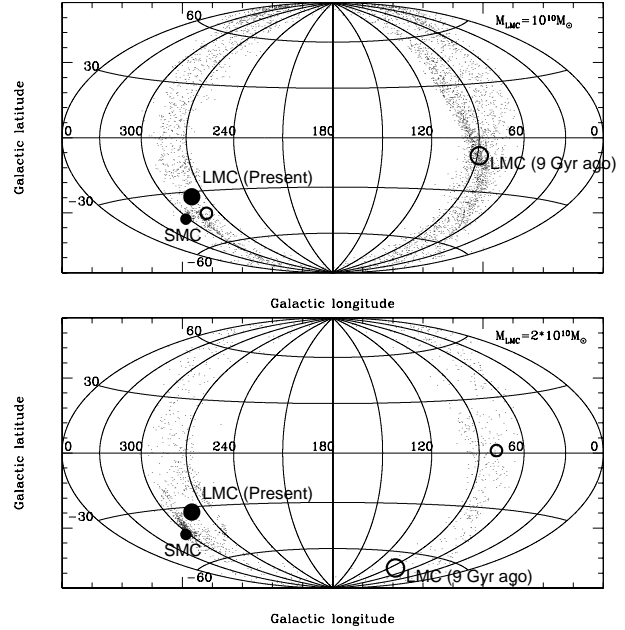,width=8.cm}
\caption{
The final distribution of old stars stripped from the
LMC disk in an Aitoff projection for the fiducial model (upper)
and for the model 9 (lower).
The preset locations ($T$ = Gyr) of the LMC and the SMC
are indicated by a large filled circle and a small filled one,
respectively, whereas the locations
of the LMC and the SMC $\sim$ 9 Gyr ago ($T$ = $-8.8$ Gyr)
are indicated by a large open circle and a small open one,
respectively.
}
\label{Figure. 24}
\end{figure}

\subsection{Dependences on model parameters}

We have described the results of the fiducial model
in which the model parameters are consistent with
the latest observations and thus the most realistic
and reasonable. However, there are still some observational uncertainties
in estimating the total mass of the LMC (vdMAHS). 
Considering this uncertainty, 
we mainly  show the results of the more massive model (model 9) with
larger LMC mass ($M_{\rm LMC}$ = 2.0 $\times$ $10^{10}$ $M_{\odot}$)
in this subsection.
We also briefly summarize  the parameter dependences  
on less important parameters (e.g., orbits of the Clouds and $f_{\rm r}$).
 
\subsubsection{LMC mass}

The adopted LMC mass $M_{\rm LMC}$ is consistent with
the observation by vdMAHS but
smaller than that adopted in most of previous studies
with $M_{\rm LMC}$ = 2.0 $\times$ $10^{10}$ $M_{\odot}$ (GN).
Accordingly, it is important to investigate how
the LMC evolution depend on its mass if it has a larger mass. 
The stellar disk of  the more massive LMC in the model 9 (10)
is less susceptible to tidal perturbation
from the Galaxy and the SMC than that in the fiducial model,
because it is  more strongly self-gravitating.
As a result of this, the SMC in this model does not so strongly influence
the LMC disk dynamically even after it dynamically couples with
the LMC ($T$ $\sim$ $-6$ Gyr) in comparison with the fiducial model. 
We summarize our principle results of the more massive LMC model 9  as follows.

(1) The formation of a global stellar bar cannot be completed
until $T$ = $-0.5$ Gyr. This delayed formation of the stellar 
bar is due to the stronger self-gravity
which prevents the Galactic tidal field from exciting
the non-axisymmetric bar instability  in the LMC disk.
Only spiral arms are excited by the Galactic tidal field
in the early dynamical evolution of the LMC.
As a natural result of the delayed  
bar formation, the bending/warping associated with
the bar formation occurs around $T$ = 0 Gyr
and is accordingly seen in the edge-on view of the LMC at 
$T$ = 0 Gyr (See Figure 22).

(2) As shown in Figure 22, the disk cannot be thickened so
much compared with the fiducial model owing to the relatively
weak tidal perturbation from the Galaxy.  
Figure 23 shows that the vertical velocity
dispersion (${\sigma}_{\rm z}$) consequently increases very slowly during 
the Galaxy-LMC-SMC interaction until the last collision between
the LMC and the SMC at $T$ = $-0.2$ Gyr.
The final  ${\sigma}_{\rm z}$ is $\sim$ 2.2 times
larger than the initial value and thus  is  a factor of 1.4
smaller than that of the fiducial model with a smaller $M_{\rm LMC}$  
($10^{10}$ $M_{\odot}$). This result implies that 
kinematical evolution of the stellar disk of the LMC
strongly depends on its mass during the Galaxy-LMC-SMC interaction.

(3) The abrupt increases of ${\sigma}_{\rm z}$ between $-0.2$ $<$ $T$ $<$ 0 Gyr 
seen in Figure 23 is due to the strong tidal perturbation
caused by the LMC-SMC collision. 
There are no significant differences
in the evolution of the half-mass radius ($R_{\rm h}$) between the isolated LMC 
model 10 and the model 9,  because radial mass transfer, which
drives a dramatic change of $R_{\rm h}$, is not efficient
both in the isolated model 10 and the model 9.

(4) The total mass of old stars stripped from the LMC disk
is smaller in the more massive LMC model (model 9) than
in the fiducial model, because the LMC is more strongly bounded by
self-gravity in the more massive LMC model.
As a result of this,  
the ``tidal stream'' composed of the stripped stars
shown in Figure 24 is less clearly seen in the   
more massive model.
This result implies that projected number density of the Galactic halo stars
along the possible tidal stream formed from stars stripped from
the LMC stellar disk provides a valuable information
of the LMC mass.

(5) The formation efficiency of new field stars (estimated from
the fraction of the mass of new stars to the initial gas  mass)
is similar between the more massive LMC model and the fiducial
model (43 \% and 47 \%, respectively). All of the new GCs 
are formed only  after $T$ $<$ $-3$ Gyr and most of the
newly formed GCs have the ages of $<$ 2 Gyr in the model 9.

\subsubsection{Other parameters}

The parameter dependences of the present simulations are 
summarized as follows.

(1) Final structural and morphological properties of the LMC disk
and star formation histories of field stars and GCs do not depend
strongly on $\theta$ and $\phi$ (i.e., inclination of the disk)
{\it for a reasonable set of these values}.
For example, mass fraction of gas converted into new field stars
range from 40 \% to 47 \% for the models 1, 3, 4, 5, 8 and
all these models show a central stellar bar composed both of old
field stars and new ones at $T$ =0 Gyr.

(2) The present results do not depend on the shape of the dark matter
halo (models 5 and 6), essentially because the model parameters
of the NFW halo are chosen such that $V_{\rm m}$ and the shape
of the rotation curve are  
similar to those  of the fiducial model. 
The epoch of the commencement of the GC formation
does  not depend on the shape of the dark matter halo,
which confirms that the most recent episode of globular
cluster formation in the LMC is related to the commencement of
strong tidal interactions between the LMC, the SMC and the Galaxy.

(3) The star formation histories of field stars and GCs 
depend on the orbital evolution of the Clouds
(model 11 $-$ 17), though the final barred morphology is
not so different between the models.
For example,  
the formation rate of field stars before $T$ = $-4$ Gyr 
is significantly higher than that after $T$ = $-4$ Gyr,
for the model with the orbital type I (i.e., model 15), 
in which the LMC-SMC distance does not become so small compared
with the fiducial model.
As a result of this, the age distribution of new field stars
in this model 15 has no peak around $T$ = $-2$ Gyr where the fiducial model
shows its secondary peak in the age distribution:
The age distribution of field stars provides a fossil record
of the Galaxy-LMC-SMC interaction history that is totally determined
by the orbits of these three.

(4) The model without the SMC's tidal effect (model 17) shows 
much less efficient formation of field star and GCs compared
with the fiducial model with the SMC.  For example,
the mass fraction of gas converted into new field stars
(new GCs) for the last $\sim$ 9 Gyr in the model
without the SMC is only 62 (17) \% of the fiducial model with the SMC. 
This result strongly suggests that the tidal perturbation from the SMC
plays an important role  in the star formation history of the LMC, 
in particular, in the GC formation.

(5) The dissipative dynamics of interstellar medium during
the Galaxy-LMC-SMC interaction also controls the formation
histories of field stars and GCs, though it cannot significantly
change the age distributions of these stellar populations
(models 1 and 18 $-$ 21 with different $f_{\rm r}$).
In particular,
the formation efficiency of GCs (i.e.,
the total number of new GCs) strongly depends on $f_{\rm r}$
in the sense that the efficiency is higher in the model
with larger $f_{\rm r}$ (i.e., more dissipative).

\section{Discussion}

\subsection{Born as a pair or different entities ?}

It remains unclear whether 
the Clouds were born initially as a primordial pair of galaxies 
at the epoch of galaxy formation 
(hereafter referred to as ``the primordial binary galaxy model''
just for convenience)
or formed as different entities at different
places and have only recently become a close pair for the first time
(``the recent coupling model''). 
This question is the central core of any problems related to 
the long-term evolution
of the Clouds ($\sim$  10 Gyr),
because strong  dynamical interaction between the Clouds
significantly influences not only star formation histories
but also structure and kinematics of the Clouds
if they are strongly coupled.
The primordial binary galaxy model has the following two advantages in 
explaining some physical properties of the Clouds.
Firstly, previous numerical studies showed that structural and kinematical
properties of the Magellanic stream are self-consistently explained by
the models for which the Magellanic Clouds have remained in a binary state
for the past 15 Gyr (MF; GSF; GN; YN).
Secondly, such a primordial binary galaxy model of the Clouds 
also naturally explains 
active star formation  about 0.2 Gyr ago in the LMC disk (GSF) and
structural properties of the stellar halo
and the very recent star formation history of the SMC (GN; YN).
MF discussed whether the binary Clouds
could be formed from a single protogalaxy via tidal fission $\sim$ 15 Gyr ago. 
Since these previous studies  showed only binary galaxy models that explain 
the above observational properties of the Cloud and the Magellanic Stream, 
it is not so clear  whether other models for which the duration
of the binary state of the Clouds does {\it not} exceed $\sim$ 15 Gyr
can equally explain physical properties.

In calculating the orbital evolution of the Clouds,
the above previous studies adopted the following two assumptions
that are not so consistent with the latest observations of LMC's structure
and kinematics (e.g., vdMAHS): 
(1) the LMC has a mass of 2 $\times$ $10^{10} M_{\odot}$ 
and (2) dynamical friction between the Clouds is negligible.
The present study has demonstrated that if we adopt more reasonable
assumptions as to the above (1) and (2),
it is not possible  for the Clouds to remain in a binary state
for the last  13 Gyr.
It has therefore suggested that the LMC and the SMC could have dynamically 
coupled relatively recently ($\sim$ 4 Gyr ago) for the first time 
and thus were born not
as a binary but as different entities. 
These results never mean that the primordial binary galaxy models (e.g., MF)
can be ruled out, because there are still some uncertainties
in the mass estimation of the LMC by vdMAHS.
It rather suggests that it depends on the model parameters.
In particular, the LMC mass determines 
how long  the Clouds can keep their  
binary state in their dynamical history:    
More precise mass estimation of the LMC is doubtlessly worthwhile
for the better understanding of the orbital evolution of the Clouds.

Formation histories of field stars and globular clusters and
structural and kinematical properties in the LMC are observed to be different
from those in the SMC (e.g., van den Bergh 1981; 2000).
For example, the ``age gap'' 
(i.e., only one cluster with the age ranging from $\sim$ 13 Gyr
to 3 Gyr) observed in LMC's globular clusters 
does not   exist in SMC's globular clusters (e.g., Piatti et al. 2002).
Since strong tidal perturbation triggers the formation of
globular clusters (e.g., Bekki et al. 2002),
it is not clear why formation histories of star clusters are different between the two
in the primordial binary model,
in which  the Clouds have been  perturbing 
one another from their initial  dynamical state until now. 
The observational fact that the age gap is seen
only in the LMC can be understood in terms of the
recent coupling model as follows.
The LMC was formed as
a relatively low surface brightness galaxy, being more distant
($\sim 150$\,kpc, corresponding to the apocenter of its early orbit)
from the Galaxy so that the Galactic tidal field alone could not trigger
cluster formation efficiently until it first encounters with the SMC.
In contrast, the less massive SMC, which is therefore more susceptible to
the Galactic tide, was born less distant ($\sim 100$\,kpc) from the
Galaxy, and thus influenced by the Galaxy strongly enough to form
globular clusters from the early evolutionary stage (several to 10\,Gyrs
ago). Thus  the difference in cluster formation histories between the Clouds
can be due to  the differences in
the birthplaces and initial masses between the two
in the recent coupling model.

Although Magellanic-type galaxies that appear to be pairs of galaxies
are not rare (e.g., Freeman 1984),
it is not so observationally clear whether these apparently pairs of galaxies
were formed {\it initially} as pairs at the epoch of galaxy formation 
or have {\it only recently} become pairs
owing to recent tidal capture or interaction of galaxies (e.g., Helou 1984).
It is accordingly difficult for us to
derive some hints on the above question as to the binary
state of the Clouds  
from observations on other Magellanic-type pair.
Currently available proper motion data for the Clouds
has the accuracy of $\sim$ 3 mas yr$^{-1}$ 
(corresponding to $\sim$ 700 km s$^{-1}$ error of  a single star
for the distance to the Clouds)
so that we cannot determine 
the duration of the LMC-SMC binary state 
directly from observational data on the proper motion
and the radial velocities of the Clouds (e.g., Westerlund 1997).
Future astrometric missions with the $\sim$ 10 $\mu$as accuracy
(corresponding to $\sim$ a few  km s$^{-1}$)
will allow us to derive an unambiguous answer for the problem
as to the binary state of  the Clouds.

\subsection{Stellar halo formation in the LMC and Magellanic-type dwarfs}

Old, metal-poor  stellar halo populations of a galaxy have long been
considered to be ``fossil records'' which contain
valuable information on  dynamical and chemical evolution 
of the galaxy (e.g., Eggen, Lynden-Bell, \& Sandage 1962).
Recent numerical simulations 
based on the currently favored cold dark matter (CDM) theory of galaxy
formation have demonstrated that 
basic physical processes involved in the formation of the stellar halo
are described by both dissipative and dissipationless merging of subgalactic
clumps and their resultant tidal disruption in the course
of gravitational contraction of the Galaxy at high redshift (Bekki \& Chiba 2000, 2001).
Previous studies also suggested that  (1) merging of subgalactic clumps
is an  essentially important process for the stellar halo formation
in disk galaxies like the Galaxy and (2) the physical properties
of the halos depend on the details of the merging processes of subgalactic clumps
(Bekki \& Chiba 2000, 2001).
It is, however, unclear 
(1) whether stellar halos exist in less luminous disk (or irregular/dwarf) galaxies
like the LMC and
(2) how they are formed in the course of their  formation.

The present study has demonstrated that (1) the stellar halo is formed 
from redistribution (in space) of stars initially within the outer 
part of the disk in the LMC {\it even if the LMC has initially no stellar halo},
(2) the developed stellar halo contains some fraction of younger stars,
and (3) the spatial distribution of the outer stellar halo 
is not homogeneous.
Accordingly the present study suggests that 
the formation process of the LMC's stellar halo 
differs from the Galactic one
in that it is not associated with any merging of subgalactic clumps.
The above result (1) furthermore implies that
{\it (i)}  the LMC could either have no or little amount of 
stellar mass in its halo at the epoch of its formation,
even if it is now observed to have the kinematically hot
stellar  halo (Minniti et al. 2003; Alves 2004),
and {\it (ii)} it acquired the substantial mass of the halo
owing to the strong tidal interaction between
the Galaxy and the SMC (This could be true for the SMC, which could suffer
more severe tidal perturbation from the Galaxy and the LMC than the LMC
could).
The present results thus raise the following two issues as to the
stellar halo formation in less luminous disk galaxies:
(1) whether less luminous disk galaxies are formed initially without
old stellar halos and 
(2) if so, why no stellar halos are formed in such galaxies
at the epoch of galaxy formation.

Recently the physical properties of old stellar halos in such less luminous
galaxies have been investigated, in particular, in dwarf galaxies in
the Local Group (e.g., Demers et al. 2003).
One of the intriguing results is that NGC 3109, a Magellanic-type dwarf
on the outskirts of the Local Group, contains carbon stars nearly exclusively
in and near its disk component, whereas NGC 6822, a galaxy with the same
morphological type, has an extended intermediate-age halo as well as an old halo.
This suggests that the formation of an old stellar halo is strongly suppressed
in some of Magellanic-type dwarfs like NGC 3109.
The following two mechanisms are considered for suppressing the formation of
a stellar halo.
One is that while the formation of disk galaxies like NGC 3109 may accompany
merging/accretion incidents of subgalactic clumps, if such clumps were totally
gaseous without containing old stars, the accretion/merging of gaseous clumps
would not leave a diffuse halo component, which originates from tidal disruption
of pre-existing stars.
The other is that the initial gas distribution in these galaxies were so diffuse
that star formation could happen only after the settlement of gas
onto the disk plane, whereby the regions with high gas density emerged.

If we adopt the currently favored theory based on hierarchical assembly of
CDM (White \& Rees 1978), the above second scenario seems less likely,
because the CDM model predicts higher overdensities for galaxies embedded in
less massive dark halos. Then, if the first scenario is the case,
what is the most likely mechanism for the suppression of star formation
within subgalactic clumps which end up with less luminous
(Magellanic-type dwarf) disk galaxies?
One possible mechanism is that thermal and/or kinematic feedback supplied by
supernovae significantly suppresses star formation in subgalactic clumps which
end up with less luminous disk galaxies (e.g., Dekel \& Silk 1986).
In this mechanism, less massive clumps are susceptible to supernova explosions
owing to their shallower gravitational potential. Alternatively, the UV
background in the Universe suppresses the formation of dwarf galaxies via
photoionization effect (e.g., Bullock, Kravtsov \& Weinberg 2000), in such
a way that less massive galaxies with lower virial temperature may be more
affected. We note that these mechanisms are also clues to solving the problem of
overabundance in the number of CDM subhalos in the Local Group (Klypin et al. 1999).
Better understanding of such suppression effects of galaxy formation may also
resolve the current issue of stellar halos in less luminous disk galaxies.

The present study predicts  that if the major component
of the  stellar halo in  a less luminous (Magellanic-type dwarf) 
disk galaxy is formed by tidal interaction with other more luminous galaxies,
the disk galaxy has (1) relatively young  stellar halo populations,
most of which come from the outer disk stars formed {\it before} the interaction
and (2) the disk has a thick disk formed thorough the disk heating by tidal interaction.
Therefore, future observations on (1) age and metallicity distributions of halo stars
in  Magellanic-type dwarf disk galaxies and (2) statistical correlations between
the presence of the outer stellar halos and that of the faint thick disk components
will help us to determine whether the stellar halos in such dwarfs
are formed from tidal galaxy interaction rather than from primordial
merging/accretion of subgalactic clumps.
Heidmann et al. (1972) revealed that the intrinsic flattening in disk galaxies
decreases (i.e., less flattened)  abruptly from Sm to Im Hubble types.
If this less flattened nature is due to disk heating in these dwarf irregulars, 
it is an observationally interesting question (related to the stellar halo
formation via tidal interaction)
whether spherical stellar halos are more likely to
be observed in Im rather than Sd galaxies.
In the stellar halo formation scenario via tidal interaction,
the age distribution of  halo stars  in a galaxy
depends strongly on when the galaxy interacted with other galaxy. 
Therefore, future observations on the age distribution
of halo stars will also provide valuable information
on the past interaction history of the galaxy.

\subsection{Origin of the LMC's stellar bar}

Recently several observational studies have attempted to derive age and metallicity  
distributions of stellar populations in the bar region of the LMC
in order to constrain the star formation history (SFH) in the bar
(Elson et al. 1997; Ardelberg et al. 1997; Holzman et al. 1999; Olsen 1999;
Smecker-Hane et al. 2002).
There however exists some discrepancy in the results of the SFH of the bar between
different observations, possibly because 
authors investigated SFHs of 
different regions within the bar using different number of stars
analyzed (e.g., Smecker-Hane et al. 2002). 
For example, Elson et al. (1997) investigated photometric
properties of $\sim$ 15800 stars obtained by the $HST$ 
for the inner LMC disk and 
found a possible evidence of
a later starburst around 1 Gyr ago which may be responsible for the bar formation
in the LMC disk.
Smecker-Hane et al. (2002) revealed that 
star formation of the dominant populations in the LMC bar
occurred from 4 to 6 and 1 to 2 Gyr ago.
Olsen (1999) suggested that the LMC bar region appears to have high levels of  
star formation activity as long as $5-8$ Gyr ago: The LMC bar is dominated by
old stellar populations.

The present study has demonstrated that (1) a large fraction of stars 
(up to 50 \%) in the bar are formed during the strong tidal interaction
between the Clouds and the Galaxy and have relatively younger ages
and (2) there can exist a steep  age gradient of stellar populations
along with and perpendicular to the bar in the sense that
the outer regions of the bar contain only a smaller fraction of young
stars.
We therefore suggest that 
the formation of the observed young stellar populations in the LMC bar region 
(e.g., Elson et al. 1997; Smecker-Hane et al. 2002)
is closely associated with the efficient star formation
within  the bar for the last  several Gyr (in particular, $\sim$ 2 Gyr ago).
We also suggest that the observed discrepancy in the SFH of the bar
could be due partly to the radial gradient of ages of stellar populations
within the bar: The mass fraction of young stars and the mean age of 
stellar populations for a target field of the LMC bar 
in previous observations depend strongly
on the distance of the field from the center of the LMC.

The present results imply that the LMC bar was formed not  {\it spontaneously}
from global bar instability  in the early evolution stage of the LMC
but from {\it tidal perturbation} by the Galaxy and the SMC.
Several numerical studies have already shown that stellar bar can be formed
via tidal interaction of disk galaxies for variously different parameters of
galaxy interaction 
(e.g., Noguchi 1987; Byrd \& Valtonen 1990). 
The mass fraction of stars within a disk embedded by a massive dark matter
halo must be at least larger than $0.4-0.5$  so that the bar is spontaneously
formed from global bar instability (e.g., Sellwood \& Carlberg 1984).
The total visible mass of stars of the LMC 
with $L_{V}$ = 3.0 $\times$ $10^9$ $L_{\odot}$
is $\sim$ 2.7 $\times$ $10^9$ $M_{\odot}$
for  $M/L_{V}$ = 0.9 $\pm$ 0.2) 
whereas the total dynamical mass of the LMC
is (8.7 $\pm$ 4.3) $\times$ $10^9$ $M_{\odot}$ within 8.9 kpc
(vdMAHS).
Therefore, the spontaneous bar formation in the LMC disk is not
likely to occur and thus the bar is likely to have formed relatively
recently from  external tidal perturbation.
Using numerical simulations, 
Noguchi (1996) demonstrated that stellar bars formed from external tidal perturbation
(``tidal bars'')
have a relatively flat density profile along the major axis of the bars
with ``shoulders'' (abrupt steepening of the gradient) at the bar ends.
We thus suggest that future observational studies on 
the radial density profile {\it  for the young stellar populations} of the LMC   
confirm the bar's radial profiles characteristic of the tidal bars,
if the LMC bar was formed from tidal interaction with the Galaxy and the SMC
relatively recently.

\subsection{The Age gap problem}

Precise estimation  of an age of each individual star cluster
in the LMC leads to the determination of the cluster age distribution
and thus to the better understanding
of the star formation history of the LMC 
(e.g., Searle et al. 1980; Hodge 1983, 1988; Mateo 1988).
Differences in spatial distributions between clusters with
different ages provide some information on the spatial
variation of the star formation history of the LMC (e.g., van den Bergh 1981).  
The age distribution of the LMC clusters shows a gap extending
from 13 to 3 Gyr with only one cluster (ESO 121-SC03) within this gap,
which is not seen in the SMC clusters 
(Jensen et al. 1988; 
Da Costa 1991; Geisler et al. 1997;
Rich et al. 2001;  Piatti et al. 2002).
The following three  possible scenarios are proposed for
explaining the above ``age gap'' in the LMC clusters.
First is that  star cluster formation 
after the initial formation of old globular clusters $\sim$ 13 Gyr ago
(at the epoch of the LMC formation)
had been suspended until very recently  $\sim$ 3 Gyr ago. 
The second scenario is 
that although cluster formation has ceaselessly continued
until now, only star clusters with ages ranging from 13 to 3 Gyr are tidally
stripped.
Third is that  star clusters with ages  between 3 and 13 Gyr were preferentially
destroyed by the LMC tidal field to become field stars.

For the above second scenario to be viable,
the star clusters with ages between 3 and 13 Gyr should be formed
preferentially in the outer LMC's halo, 
where tidal stripping of the clusters by the Galaxy
is very efficient in this scenario.
Given the possible observational evidence  that
both young ($<$ 3 Gyr old) and old ($\sim$ 13 Gyr) clusters
show disky distributions (e.g., Schommer et al. 1991; van den Bergh 2000),
it is unclear why only clusters with ages between 3 and 13 Gyr 
are formed in the outer halo region of the LMC.
Therefore the second scenario
is regarded as rather less likely one.
Regarding the third scenario,
it could be  possible for clusters with ages between 3 and 13 Gyr 
to be preferentially destroyed,
only if they have typically lower densities and masses compared with
other LMC clusters.
Since no previous theoretical models predicted age dependences of structural
properties of globular clusters (e.g., Harris 1991 and reference therein),
the third scenario is  equally less likely one.

Thus, if the first scenario is only a reasonable one,
the essence of the problem related to the above ``age gap'' in the LMC clusters
is what mechanism is responsible for the abrupt reactivation of cluster formation
in the LMC $\sim$  3 Gyr.
Using numerical simulations, Bekki et al. (2004b) first discussed 
this problem in the context of mutual tidal interaction between the Clouds
and the Galaxy. 
Bekki et al. (2004b) and the present study 
have proposed that the epoch of reactivation of cluster formation
corresponds to the commencement of strong tidal interaction between
the LMC and the SMC, which disturbs the LMC gas disk,
enhances cloud-cloud collision rate,  and consequently triggers 
cluster formation. 
In this scenario,  the tidal interaction between the LMC and the Galaxy
alone cannot increase so dramatically the number of cloud-cloud collisions
leading to cluster formation between 13 and 3 Gyr ago.

The above scenario could be just one of promising scenarios explaining 
the origin of the age gap, we accordingly point out
two possible alternative scenarios below.
Byrd et al. (1994) numerically investigated the orbital evolution
of the Clouds and Leo I and thereby 
proposed a scenario  that the Cloud left M31 $\sim$ 10 Gyr ago
and were tidally captured by the Galaxy several  Gyr ago.
Although they did not discuss their results in terms of
the age gap of the LMC clusters,
it is not unreasonable to expect  that
the cluster formation could be suddenly triggered
by the strong Galactic tidal force
when the LMC first experienced 
the pericenter passage with respect to the Galaxy.
One of the alternative scenarios is thus that
the origin of the age gap is closely associated
with the first pericenter passage of the LMC that once belonged to M31. 
In this scenario, 
Byrd et al. (1994) showed that the epoch of the first pericenter passage is 
ranging from 4.6 Gyr to 12.2  Gyr ago for a relatively narrow parameter space
of the orbital evolution.
Therefore it remains less clear whether  
LMC's first pericenter passage  is 
{\it not} $\sim$ 6 Gyr ago
but $\sim$ 3 Gyr ago. 
Numerical studies
with more variously different yet reasonable initial orbital parameters
of the LMC and with more realistic mass models of the Local Group
will thus assess the viability of this scenario.

The other alternative scenario is that 
$\sim$ 3 Gyr ago corresponds to the epoch when
the LMC's disk gas begins to interact with the Galactic
halo plasma and form clusters owing to bow-shocked induced
star formation in the LMC disk.
de Boer et al. (1998) pointed out that 
a shock induced by the ram pressure of the halo plasma
can induce star formation in the LMC
for a reasonable set of parameters of halo gaseous  density and temperature
and the relative velocity of the LMC with respect to the Galaxy.
Hydrodynamical simulations demonstrated that
ram pressure  of
the intracluster/intragroup medium  strongly compresses a self-gravitating  gas cloud
within a short time scale ($\sim$ $10^{7}$ yr), dramatically
increasing the central gas density,
and consequently causing efficient  formation of 
a compact star cluster within the cloud (Bekki \& Couch 2003).
Although these two works suggest that cluster formation via ram pressure of
the Galactic halo plasma is possible,
it is not clear why cluster formation via ram pressure becomes possible 
for the first time only $\sim$ 3 Gyr ago,
given the fact that the LMC had experienced several pericenter passages before
3 Gyr ago.

One of the possible reason for the sudden interaction between the LMC gas
and the Galactic halo gas $\sim$ 3 Gyr ago is that 
dynamical friction from the Galactic dark matter halo 
causes the orbital decay of the LMC so slowly that  
the LMC cannot  approach so closely the outer edge of the  Galactic hot plasma
until recently ($\sim$ 3 Gyr ago).
There could be more reasonable scenarios for the age gap problem
other than the three discussed above.
The age gap reflects the complicated interaction history between
the Clouds and the Galaxy,
which depends almost exclusively the details of the orbital evolution
of the Clouds. 
This emphasizes the importance of proper motion measurement with
the accuracy of an order of $\mu$as for the Clouds in
clarifying the origin of the age gap of the LMC clusters.

\section{Conclusions}

We have performed numerical simulations for the dynamical and chemical 
evolution  of the LMC interacting with the Galaxy and the SMC for
the last 9 Gyr. 
The main results are summarized as follows.

(1) Tidal interaction between the Clouds and the Galaxy plays a major role
not only in the morphological transformation  of the LMC disk but also
in the formation history of field stars and star clusters.
The interaction transforms the initially thin, non-barred LMC disk
into the three different components; the thick disk, bar, and
kinematically hot stellar halo.
The central bar formed during the tidal interaction is composed both of
old field stars and newly formed ones with each fraction being equal
in the innermost part.
The final thick disk has the velocity dispersion 
of $\sim$  30 km s$^{-1}$ and shows rotationally supported kinematics 
with $V_{\rm m}/{\sigma}_{0}$ $\sim$  2.3.
The outer stellar  disk ($R_{\rm L}$ $>$ 5 kpc) surrounding the
central bar is highly elliptic resulting from the tidal interaction
between the Galaxy and the SMC.

(2) The stellar halo is formed during the interaction
as a result of redistribution of stars initially within
the outer part of the thin LMC disk.
The stellar halo is thus composed mainly of old stars originating 
from the outer part of the initially thin LMC disk.  
The outer halo shows velocity dispersion of $\sim$ 40 km s$^{-1}$ 
at the distance of 7.5 kpc from the LMC center 
and has somewhat inhomogeneous distribution of stars. 
The stellar halo contains  relatively young, metal-rich stars
with the mass fraction of 2 \% in the halo.
Therefore the MDF of the halo is determined by the old stars
and shows a peak around [Fe/H] $\sim$ $-1.3$ for a reasonable 
set of parameters of the initial MDF in the stellar disk of the LMC.

(3) Star formation rate in the LMC disk is moderately and repeatedly enhanced 
owing to the repetitive interaction between the Clouds and the Galaxy.
The star formation rate increases from  $\sim$ 0.1 $M_{\odot}$ yr$^{-1}$
to $\sim$ 0.4 $M_{\odot}$ yr$^{-1}$ at the first pericenter passage of 
the LMC with respect to the Galaxy about 7  Gyr ago.
The star formation rate also becomes moderately high
when the Cloud begins to interact violently 
(with the LMC-SMC pericenter of less than 10 kpc)
about $\sim$ 3.5 and 2 Gyr ago.
Most of the new stars ($\sim$ 90 \%) are formed within the
central 3 kpc, in particular, within the bar for the last 9 Gyr. 
Consequently, the half mass radius is different by a factor
of 2.3 between old field stars and newly formed ones.
These structural differences between field stars with difference ages
are characteristic of the LMC disk under tidal interaction with the Galaxy
and the SMC. 

(4) Efficient GC formation does not occur until the LMC starts interacting
violently and closely with the SMC ($\sim$ 4 Gyrs ago).
This is due to the fact that cloud-cloud collisions
with moderately high relative speed (30 $\le$ $V_{\rm rel}$ $\le$ 100 km s$^{-1}$)
and with small impact parameter ($b$ $<$ 0.25) required for GC formation
occurs the most frequently when both the SMC and the Galaxy dynamically  
influence the LMC's disk strongly. 
The newly formed GC system has a disky distribution with
rotational kinematics and its mean metallicity is $\sim$ 1.2 higher
than that of new field stars because of the pre-enrichment 
by the formation of field stars prior to cluster formation.

(5) About 15(20) \% of the field stars (gas) 
initially within the LMC disk are tidally stripped
to form a great stellar (gaseous) circle of a relic stream around the Galaxy
during the last 9 Gyr evolution of the LMC.
The great stellar circle shows inhomogeneity in some parts and is composed only of
metal-poor old stars. The unique distributions of distance and radial velocity
in the tidal stream may well enable us to pick out the stream among the
Galactic halo stars.
The stellar  total mass of the tidal stream  depends on the initial mass of the LMC
(i.e., smaller for the larger LMC mass), so that the stellar number density
along the stream provides valuable information on the LMC mass.

(6) The LMC evolution depends on its initial mass and 
orbit with respect to the Galaxy and the SMC.
In particular, the epoch of the bar and the thick disk
formation is determined by the LMC mass in such a way that 
the stellar bar and the thick disk are formed later
in the model with a larger LMC mass.
The mass fraction of the stellar halo  is smaller
for the model with a larger LMC mass.
These are  essentially because the LMC with a larger mass
is more strongly bounded by its self-gravity
so that the tidal perturbation from the Galaxy and the SMC
does not so significantly influence the dynamical evolution of the LMC.

Based on these results, we have discussed the origin of the stellar halo
in less luminous late-type galaxies such as the Clouds,
the formation of LMC's stellar bar and thick disk,
the origin of the age gap of the LMC globular clusters, 
and the difference in formation histories of field stars and globular clusters
between the LMC and the SMC.
It has been also pointed out that 
future proper motion measurements of the Clouds with $\sim$ 10 $\mu$ac accuracy
(Perryman et al. 2001) to estimate their past 3D orbits
in an unprecedentedly precise manner will provide us invaluable information 
on the complicated interplay between
dynamical evolution of the LMC 
and its star formation history.

\section{Acknowledgement}
We are  grateful to the  referee Gene Byrd for valuable comments,
which contribute to improve the present paper.
K.B. acknowledges the Large Australian Research Council (ARC).
Numerical computations reported here were carried out on GRAPE
system kindly made available by the Astronomical Data Analysis
Center (ADAC) of the National Astronomical Observatory of Japan.


\appendix
\section[]{Statistics on the duration of the LMC-SMC binary status}

As suggested by  Bekki et al. (2004b) and the present study,
the epoch when the LMC and the SMC become dynamically coupled
(i.e., when the pericenter distance of the LMC-SMC orbit
is as small as 10 kpc) is a critical moment for the LMC,
because the combined tidal effect of the Galaxy and the SMC
starts influencing significantly the LMC evolution  
after the dynamical coupling.
It is thus important to investigate when they are  the most likely to
become coupled during their dynamical evolution.
Although the proper motion of the Clouds have been already 
derived by several authors (e.g., KB; vdMAHS),  the measurement error in
these observational studies is so large that  precise
estimation of current velocity components $(U_{\rm L},V_{\rm L},W_{\rm L})$
and $(U_{\rm S},V_{\rm S},W_{\rm S})$ 
within an error of $\sim$ a few km s$^{-1}$
cannot be made directly from the observations.
As shown in previous studies on the orbital evolution of
the Clouds (e.g., MF),  
only a velocity difference of $\sim$ 5 km s$^{-1}$ 
can cause a significant difference in the orbital evolution 
and thus in the binary status of the Clouds. 
Thus currently available data alone do not allow us to
make a robust conclusion on the duration of the LMC-SMC
binary status.

\begin{table}
\caption{Orbit models}
\begin{tabular}{ccccc}
orbit model &
$F_{\rm bin,1}$(\%) & 
$F_{\rm bin,2}$(\%)& 
$t_{\rm bin,1}$(Gyr)& 
$t_{\rm bin,2}$(Gyr) \\
A & 0 & $5.5\times10^{-4}$ & 4.29  & 1.15  \\
B & 0 & $1.1\times10^{-2}$ & 4.42 & 1.10  \\
C & 16.7 & $4.1\times10^{-3}$  & 6.25 & 1.23   \\
D & 43.8 & $4.6\times10^{-2}$ & 8.88 & 1.26   \\
\end{tabular}
\end{table}

\begin{figure}
\psfig{file=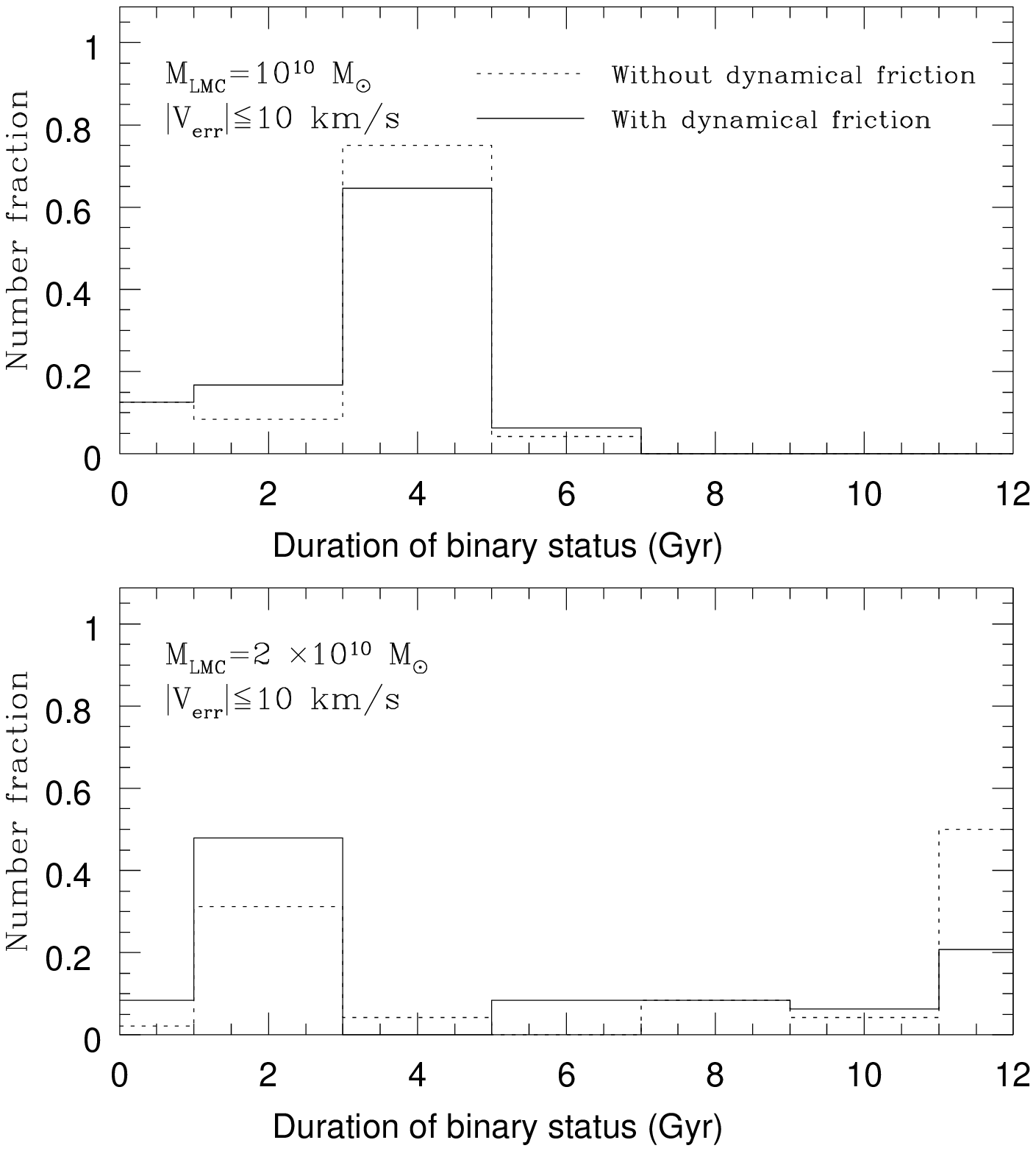,width=8.cm}
\caption{
The distributions of $t_{\rm bin}$
duration of binary status of the Clouds
for the models with $M_{\rm LMC}$ = $10^{10}$ $M_{\odot}$
(upper) and those with $M_{\rm LMC}$ = 2 $\times$
$10^{10}$ $M_{\odot}$ (lower).
$t_{\rm bin}$ represents the duration of binary status
of the Clouds and the definition of the binary status
is given in the main text.
For comparison, the results
are shown for models with (solid) and without (dotted)
dynamical friction between  the Clouds in each panel.
For convenience, the normalized number of models
is given for each bin of the LMC-SMC binary duration.
Here only the models in which each component
of the current velocities
of the Clouds ($(U_{\rm L},V_{\rm L},W_{\rm L})$ and
$(U_{\rm S},V_{\rm S},W_{\rm S})$)
is within $\pm$ 10 km s$^{-1}$ of the corresponding
component in the model.
}
\label{Figure. 25}
\end{figure}

\begin{figure}
\psfig{file=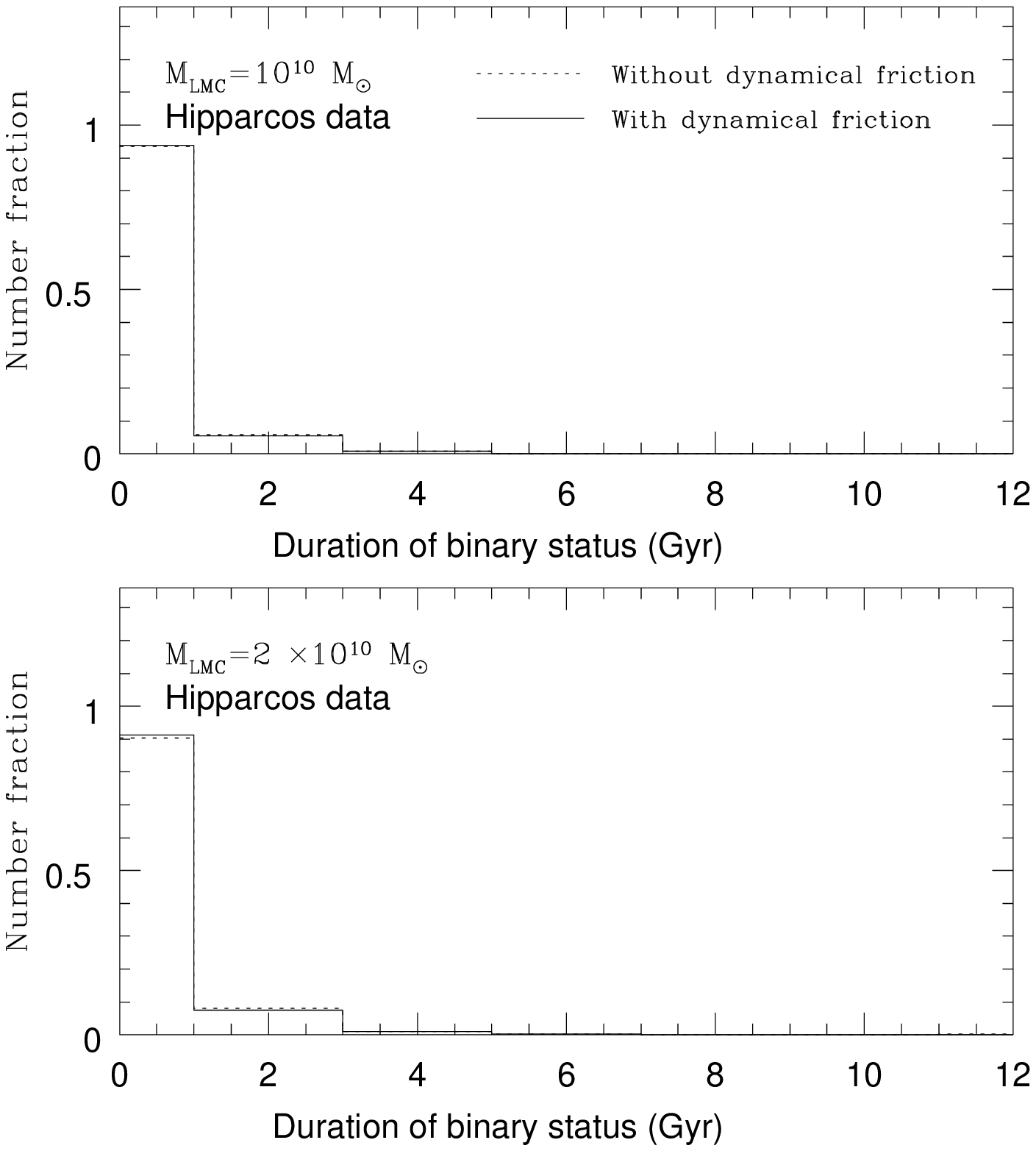,width=8.cm}
\caption{
The same as Figure A1 but for the models with the current
velocities of the Clouds similar to those
obtained from Hipparcos data (KB).
}
\label{Figure. 26}
\end{figure}

Therefore we here make a statistical argument on the duration
of the LMC-SMC binary status
by investigating every possible orbits of the Clouds for
a set of reasonable parameters of gravitational potential
for the Galaxy and the Clouds. 
Using the backward integration scheme (MF) and
the models for gravitational potential and dynamical friction
adopted in this study
for the Galaxy and the Clouds (i.e., the equation (1) - (5)),
we first calculate the orbital evolution of the Clouds for each model
with a given set of parameters for their initial velocities. 
Then we estimate the epoch when the LMC-SMC distance is for the 
first time larger than 50 kpc that is adopted by previous
numerical studies (MF; GSF)
for the threshold of the binary status.

Initial velocity components of $U_{\rm L}$ and $V_{\rm L}$
($U_{\rm S}$ and $V_{\rm S}$) for  the LMC (the SMC)
range from $-320$ km s$^{-1}$ to 320 km s$^{-1}$
in the orbital investigation.  $W_{\rm L}$ ($W_{\rm S}$) is derived 
from given $U_{\rm L}$ and $V_{\rm L}$ ($U_{\rm S}$ and $V_{\rm S}$)  
in the LMC (the SMC) so that the radial velocity of the LMC (the SMC)
is consistent with the observed radial velocity of the LMC (the SMC).
We survey the LMC-SMC binary orbit every 3.2 km s$^{-1}$
in the range of $(U_{\rm L},V_{\rm L},W_{\rm L})$-space 
($(U_{\rm S},V_{\rm S},W_{\rm S})$-space)
and thus the total number of the orbits investigated in this study
is an order of ($10^2 \times 10^2$) $\times$  
($10^2 \times 10^2$)  = $10^8$. 
Such a numerous number of orbital investigation enable us
to provide a statistical argument on the duration of the LMC-SMC
binary status.

We adopt the four different orbital models with different $M_{\rm LMC}$
and with or without dynamical friction: 
Model O1  with $M_{\rm LMC}$ = $10^{10}$ $M_{\odot}$ 
and with dynamical friction,
O2  with $M_{\rm LMC}$ =  $10^{10}$ $M_{\odot}$ 
and without dynamical friction,
O3  with $M_{\rm LMC}$ = 2 $\times$ $10^{10}$ $M_{\odot}$ 
and with dynamical friction,
and O4  with  $M_{\rm LMC}$ = 2 $\times$ $10^{10}$ $M_{\odot}$ 
and without  dynamical friction.
We investigate not only the models with $M_{\rm LMC}$ consistent with
observations (i.e., O1 and O2) but also those with $M_{\rm LMC}$
significantly larger than the observed one (vdMAHS),
because we intend to compare the present results with previous
ones (e.g., GN) in which $M_{\rm LMC}$
is assumed to be 2 $\times$ $10^{10}$ $M_{\odot}$.

Based on the above models, we search for the models which meat 
the following two requirements/conditions:
(C1) The current velocities are broadly consistent with those derived
from the latest observations, 
and (C2) the Clouds can keep its binary status 
for the Hubble time ($\sim$ 13 Gyr).
We first determine the models satisfying
the above condition C1 by selecting the models with each of the current velocities 
(e.g., $U_{\rm L}$) 
within an error of  10 km s$^{-1}$ of the observationally inferred one
among  all models.
We then search for the models satisfying the above condition C2
among those satisfying the above condition C1
based on  the backward integration of the Clouds' orbits.
In the first  selection process, 
we check whether the current velocities in each model are consistent
either with those adopted by GN
or with those by KB. 

We mainly investigate (1) number fraction of orbits ($F_{\rm bin}$) satisfying the
above two conditions among all possible ones and (2)  mean duration
of the LMC-SMC binary status ($t_{\rm bin}$)
for each of the four models, O1, O2, O3, and O4.
The number fraction of orbital models in which
the above condition C1 is satisfied and the current velocities 
are within an error of 10 km s$^{-1}$ of those by GN (KB)
is represented by $F_{\rm bin,1}$ ($F_{\rm bin,2}$). 
The mean duration of the LMC-SMC binary status
for orbits with the current velocities 
within an error of 10 km s$^{-1}$ of those by GN (KB) 
is represented by $t_{\rm bin,1}$  ($t_{\rm bin,2}$).
The Table A1 summarizes the results of the orbital investigation:
Model number (column 1), $F_{\rm bin,1}$ (2), $F_{\rm bin,2}$ (3),
$t_{\rm bin,1}$ (4), and $t_{\rm bin,2}$ (5).

Figure A1 shows the distribution of $t_{\rm bin,1}$ 
of  orbits  in which the current velocities are within
an error of 10 km s$^{-1}$ of those by GN
for the O1 $-$ O4 models.
It is clear from this figure that (1) the Clouds can keep 
their binary state for only less than  7 Gyr (i.e.,
the Clouds becomes disintegrated until $T$ = $-7$ Gyr)
in the models O1 and O2,
(2) they are the most likely to keep their binary
status for $\sim$ 4 Gyr in the models O1  and O2,
(3) $t_{\rm bin,1}$ is the most likely to be $\sim$ 2 Gyr
for the O3 model, 
and (4) the Clouds are the most likely to be able to  keep their
binary status for the Hubble time {\it only if the LMC has a large mass
of 2.0 $\times$ $10^{10}$ $M_{\odot}$  without dynamical
friction between the Clouds}. 
These results imply that the Clouds are very hard to keep
their binary status for the Hubble time if a reasonable
sets of assumptions  ($M_{\rm LMC}$ = $10^{10}$ $M_{\odot}$ and
inclusion of the dynamical friction between the Clouds)   
are made for the orbital calculations.

Figure A2 shows the distribution of $t_{\rm bin,2}$ 
for  orbital models   in which the current velocities are within
an error of 10 km s$^{-1}$ of those by KB
for the O1 $-$ O4 models.
As shown in this Figure A2,
the Clouds are the most likely to be able to
keep their binary status for less than 1 Gyr, irrespectively of
the LMC mass and whether or not the dynamical friction between the Clouds 
is included in the orbital calculations.
The more massive LMC has a higher probability of keeping the LMC-SMC
binary status (i.e., larger $F_{\rm bin,2}$)
for the Hubble time in these models with and without
dynamical friction (See the Table A1). 
The results shown in Figure A1 and A2  thus strongly suggest
that the probability of Clouds keeping their binary status
is very low. 

\end{document}